# 4-Dimensional Tracking with Ultra-Fast Silicon Detectors

Hartmut F.-W. Sadrozinski, Abraham Seiden

*SCIPP, UC Santa Cruz, CA 95064, USA*

Nicolò Cartiglia

*INFN, Torino, Italia*

# 1 Abstract

The evolution of particle detectors has always pushed the technological limit in order to provide enabling technologies to researchers in all fields of science. One archetypal example is the evolution of silicon detectors, from a system with a few channels 30 years ago, to the tens of millions of independent pixels currently used to track charged particles in all major particle physics experiments. Nowadays, silicon detectors are ubiquitous not only in research laboratories but in almost every high-tech apparatus, from portable phones to hospitals. In this contribution, we present a new direction in the evolution of silicon detectors for charge particle tracking, namely the inclusion of very accurate timing information. This enhancement of the present silicon detector paradigm is enabled by the inclusion of controlled low gain in the detector response, therefore increasing the detector output signal sufficiently to make timing measurement possible. After providing a short overview of the advantage of this new technology, we present the necessary conditions that need to be met for both sensor and readout electronics in order to achieve 4-dimensional tracking. In the last section we present the experimental results, demonstrating the validity of our research path.

## 2   Particle Detectors for Space and Time Measurements

### 2.1   4-Dimensional Tracking with Ultra-Fast Silicon Detectors

The measurement of trajectories of charged particles is ubiquitous in applications of physics to a wide variety of areas. This ranges from cosmic rays in space science, to ionized molecules in mass spectrometers and charged particles in medical treatment, all the way to research on fundamental particle interactions in nuclear and particle physics. These applications using charged particles typically require measurements of particle locations to some specified accuracy, coverage of some specified detection area by a particle detector, and ability to deal with some specified rate of incoming particle hits. However, improvements in rate capability often translate into more rapid measurements and larger amounts of data collected, providing an improvement in the capability of an apparatus. A crucial tool for making such particle measurements has been the silicon sensor [1][2]. Given its relation to silicon technology through the planar fabrication process, it has allowed very high spatial granularity while covering large areas using arrays and providing the ability to accept data at very high rates when connected to appropriate VLSI electronics. A limitation has been the signal formation process, which has traditionally limited the ability to measure the arrival time of a particle to values larger than about 0.2 nanosecond. Given the speed of light, this corresponds to a flight path uncertainty of greater than 5 cm if we use the device to measure flight distances.

The recent development of a new type of silicon sensor promises to significantly enhance the capability to measure track arrival times, leading to a dramatic improvement in the capability of silicon arrays. The goal is to simultaneously maintain the high granularity for spatial measurement and the capability for high rate data collection while making very good time measurements. In fact the time measurement requires very short duration signals allowing even larger data rates than conventional silicon sensors. An array of these detectors can cover a large detection area, like the more traditional silicon detectors, and, assuming an accuracy on the arrival time of 10 picoseconds, it would enable the measurement of the length of the flight path of relativistic particles with an uncertainty of 3 mm. Such good time measurement can, however, be used in other ways than just a time-of-flight measurement. For example, it can be used to group particles simultaneously coming from a well-defined location by their common arrival time if they are moving at relativistic speeds. This can be very useful at particle colliders if many interactions are occurring reasonably close in time but resolvable by a 10 picosecond time measurement.

## 2.2 The effect of timing information

### 2.2.1 An example: the effect of timing information at High-Luminosity LHC

To illustrate the impact of timing information we look at the situation typical of a high-energy physics collider experiment, where charged particles are detected (their presence reconstructed) by a series of position measurements (e.g. by pixel detectors) spaced over the particle trajectories. Typically, there are many particles to be detected, often from several events occurring within the time window of one beam crossing. This situation is shown in Figure 1[1] for one bunch crossing with ∼ 50 overlapping events recorded by the CMS experiment [3] in 2012. As the density of events is such that they occur in different locations, as it happens in Figure 1, traditional 3-dimensional tracking information is sufficient to reconstruct each vertex correctly.

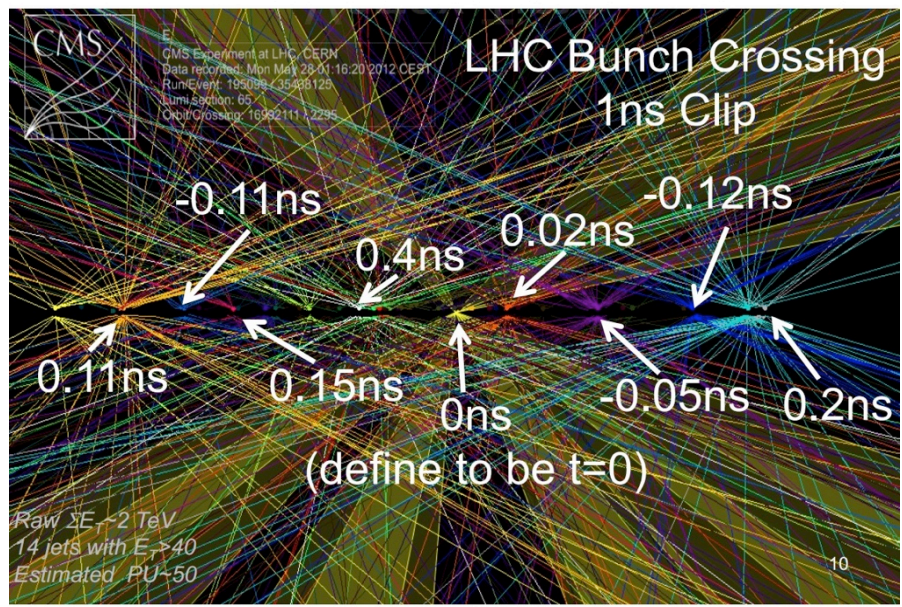

Figure 1 Interaction time of many proton-proton vertexes happening in the same bunch crossing in the case of ∼ 50 overlapping events. The vertexes are spaced 10's of pico seconds apart.

---

However this situation will substantially change at HL-LHC [4] [2] where the number of events per bunch crossing will be of the order of 150-200 and the density of events will be so large that events will be overlapping in space, as shown in Figure 2 on the left side.

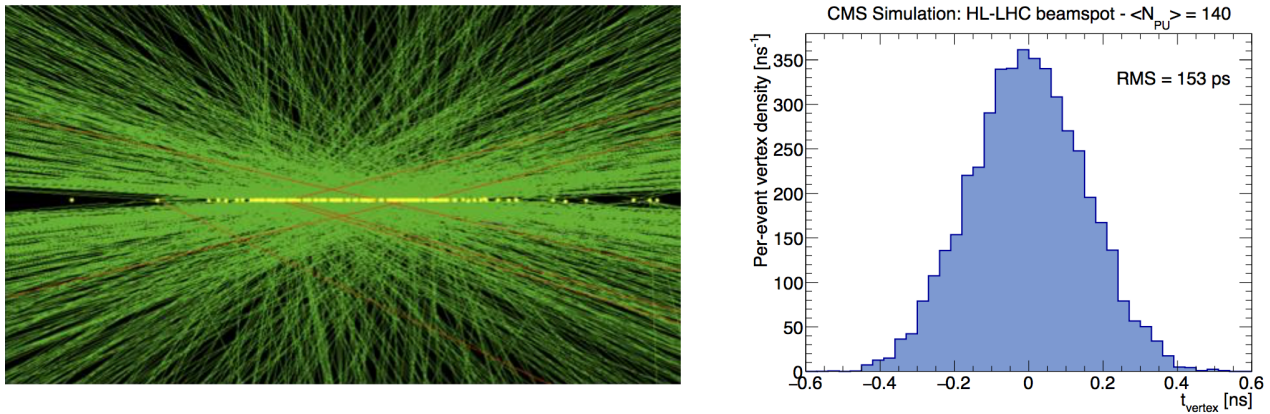

Figure 2  Left side: z-vertex distribution for a single bunch crossing at HL-LHC. Right side: the distribution of the interaction time at HL-LHC considering an average pile-up of 140 vertexes.

Assuming a vertex separation resolution from tracking of 250-300 micron along the beam direction (present resolution for CMS and ATLAS [5]), there will be 10-15% of vertexes composed by two events: this overlap will cause degradation in the precision of the reconstructed variables, and lead to loss of events.

This situation can dramatically improve with the inclusion of the timing information.  Figure 2 shows on the right side that the timing distribution of the events on the left side have an rms of ~150 ps:  in a very simplified view, a timing precision of 30 ps allows dividing the events into 5 distinct groups, each with a number of concurrent interactions equal to one fifth of the total, thus almost completely avoiding overlapping events. Timing information at HL-LHC would enable exploiting the full potential of the luminosity capability of HL-LHC and it is therefore equivalent to improved luminosity, in addition to eliminating false event assignment.

---

[2] High-Luminosity LHC is the upgrade of the current CERN LHC accelerator, scheduled to begin operation around the year 2025.

### 2.2.2   Timing information for each point or for each track

The inclusion of timing information in the structure of a recorded event has the capability of changing the way we design experiments, as this added dimension dramatically improves the reconstruction process. The most obvious simplification is that only time-compatible hits are used in the pattern recognition phase, discarding those hits that cannot be associated to a track due to an excessive time difference.

Depending on the type of sensors that will be used, timing information can be available at different stages in the reconstruction of an event. The most complete option is that timing is associated to each point of the track: in this case the electronics needs to be able to accurately measure the time of the hit in each pixel. This option is indeed quite difficult to achieve, due to the massive increase of power consumption for the readout circuits required. Nevertheless, as the potential gain that this option offers in terms of performance is the largest, we have set this option as our final goal. Figure 3 schematically shows the effect of having timing information for each hit (for additional details of fit techniques see [6]). In this specific example, a seemingly random assembly of points gets resolved into two crossing tracks and an additional random hit.

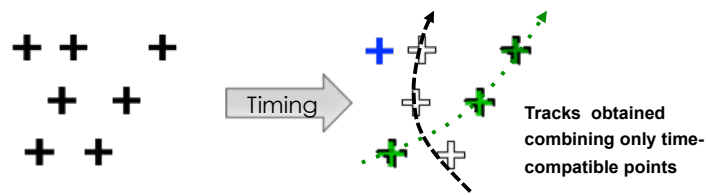

Figure 3 Effect of the inclusion of timing information in the event reconstruction at the hit level.

A very interesting option to deal with the increased power consumption mentioned above is to reduce the granularity of the timing information. To retain the full power of timing information at the event reconstruction level, i.e. to maximize the precision of the reconstructed kinematical quantities, it is actually sufficient to assign a time to each track and not to every hit. This solution is much easier than assigning time to each hit, as it can be done with a single dedicated timing layer either inside or outside the tracker volume. Figure 4 shows schematically how the time measurement helps disentangling two overlapping events: on the left side it presents the longitudinal and transverse views of tracks originating from the same point.

The addition of a 4th dimension allows distinguishing the two vertexes, so that in the presence of a timing layer the tracks can be separated into 2 events.

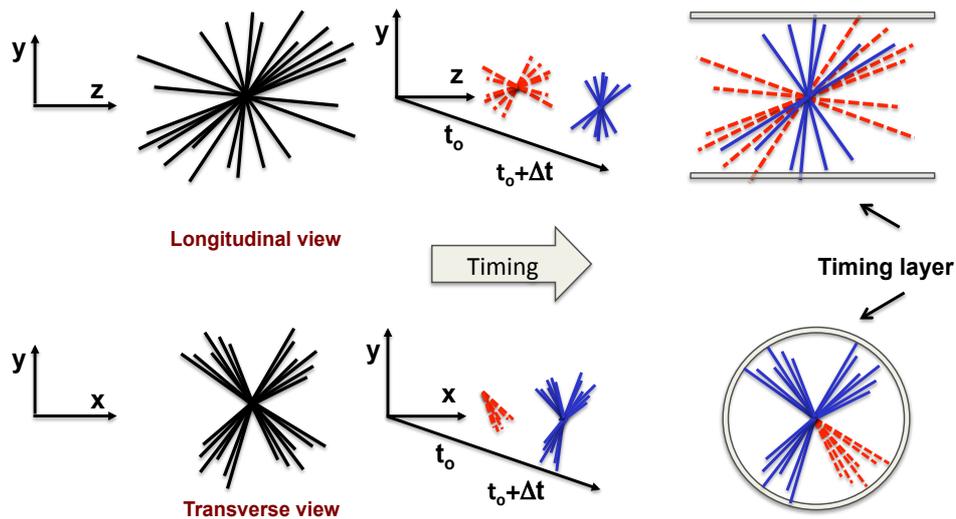

Figure 4 Schematic representation of the power of timing information in distinguishing overlapping events using a timing layer.

The timing information at the trigger level can play a key role in avoiding the saturation of the first level trigger (L1) bandwidth with fake events: as timing information can be obtained quite fast, its use in L1 decisions will allow distinguishing events with the same topology but originating in either one or many collisions, for example separating true 3-jet events from the overlap of jets from uncorrelated events or the correct identification of a forward jet as either a true jet or simply a random clustering of tracks coming from unrelated scattering.

## 2.3 Combining time and position determination: the Ultra-Fast Silicon Detector project

Having demonstrated in the preceding section that detectors with excellent timing resolution promise tangible advances in research we are turning to the question how these advantages can be implemented in existing or future detector systems to augment their capabilities. An obvious choice is to combine the timing detectors with existing detectors that afford high precision in locating particles, such as the widely used silicon detectors. Silicon detectors, as mentioned in the introduction, are based on semiconductor technology, and have properties which makes them uniquely adaptable to use in many fields of research from astrophysics to medicine and biology to nuclear and elementary particle physics: they are thin, light-weight, without need of special consumables like gases, and afford high position resolution of 10 microns or better over large area (up to hundreds of square meters). We are calling a silicon sensor with 4-dimensional imaging capabilities, i.e. high spatial granularity and added high timing resolution an "Ultra-Fast Silicon Detector" (UFSD) [7]. The UFSD development project thus benefits from the ~40 years of silicon detector R&D which explored the design, manufacturing and operations of silicon detectors for use in a variety of applications. This includes e.g. the understanding of radiation damage in ordinary silicon detectors, allowing us to concentrate on exploring those effects that are introduced by the addition of the timing capability.

### 2.3.1 Principle of operation of a silicon detector

The basic operational principles of a n-on-p silicon detector are shown in Figure 5: an external bias voltage polarizes the p-n junction inversely, creating a large depleted volume. When an incident charged particle crosses the sensor, it creates along its path electron-hole pairs (e-h). The number of e-h pairs created depends upon the particle type (for example $\alpha$-particles create many more pairs than minimum ionizing particles) and energy and on the sensor thickness [8]. The average energy loss of a charged particle in a medium as a function of the particle energy is described by the Bethe formula while the distribution of energy loss is described by a Landau curve (for a recent and complete review on silicon detector properties see [8]). For this fact, the average energy loss is 30% higher than the most probable value (MPV). In the following we will use the value MPV = 73 e-h pairs created by a minimum ionizing particle in each micron. Under the influence of the electric field, the electrons drift toward the $n^{++}$ contact, and the holes toward the $p^{++}$ contact, creating an induced current on the electrodes. The current is at its maximum, which is about 1.5 μA independent of the detector thickness (for additional details on this fact see 4.2), right after the passage of the impinging particle, and it ends when the last charge carrier reaches its electrode. Considering as an example a 300-micron thick sensor with a bias voltage of 600 V, the electron drift time is about 3 ns while the hole drift time is about 5 ns (see Table 1 for the expression of electrons and holes velocities). If the detector has segmented electrodes, the signal will be unipolar on the collecting electrode, and bi-polar on the others. In most applications the amplifier integrates the current for a given interval of time producing a signal whose amplitude is proportional to the integral of the induced current, roughly 1 fC every 100 micron of sensor thickness.

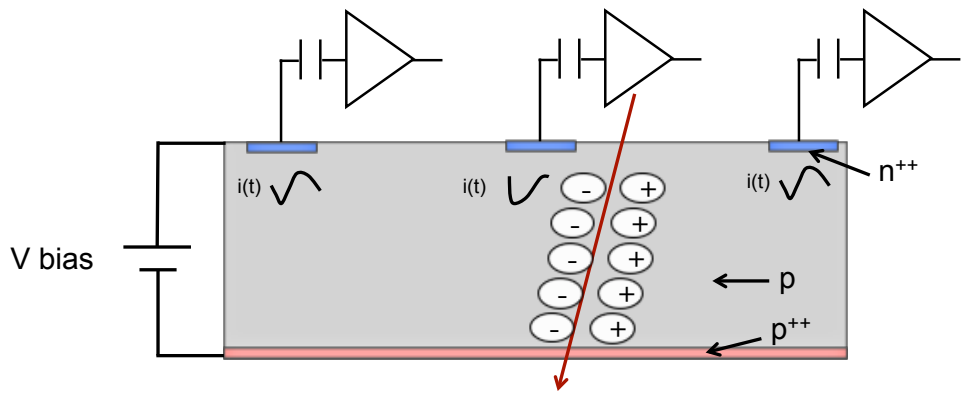

Figure 5 Basic operational principles of a silicon detector: an external bias voltage polarizes the p-n junction inversely, creating a large depleted volume. When an incident charged particle crosses the sensor, it creates electron-hole pairs whose drift generates an induced current in the electronics.

As explained in detail in the following sections, the accurate measurement of the hit time of a particle requires a large signal with a short rise time: both requirements can be fully met by thin silicon sensors with internal charge gain.

### 2.3.2 Low-Gain Avalanche Detectors (LGAD)

Charge multiplication is well understood in gases and solids and is based on the avalanche process initiated by a charge moving in large electrical fields, leading to impact ionization with a gain given by the average number of final particles created by one particle. In semiconductors this effect is used in Avalanche Photon Detectors (APD) with gain in the 100's and Silicon Photon Multipliers (SiPM) with a gain of about 10,000 [10]. In distinction to those applications, UFSD are based on Low-Gain Avalanche Detectors (LGAD) with a gain of 10-20 [11]. LGAD design is based on a modification of the doping profile where an additional doping layer of $p^+$ material (Boron or Gallium) is introduced close to the n-p junction. A simplified drawing for a traditional n-in-p (where this term indicates a n-doped electrode in a high resistivity p-doped substrate) silicon detector and that of an LGAD are shown

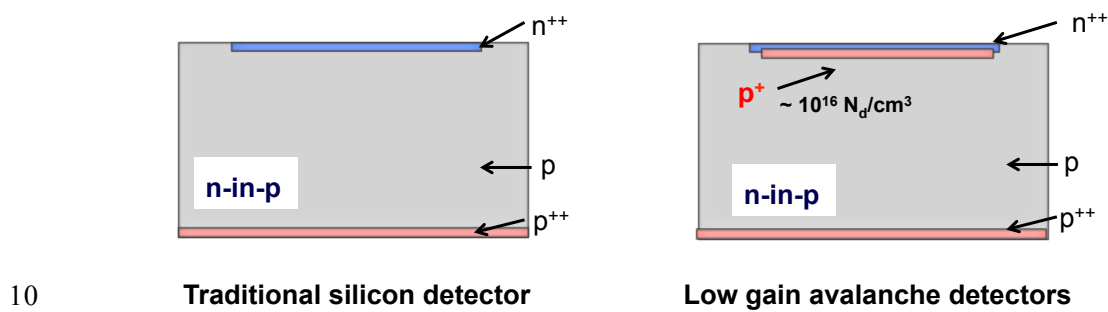

**Traditional silicon detector**          **Low gain avalanche detectors**

Figure 6 Left: no-gain n-in-p Si detector. Right: LGAD design, with the introduction of a thin $p^+$-layer below the junction. Figure taken with permission from [10].

in Figure 6.

The resulting doping profile is characterized by a large increase in doping concentration in close proximity to the junction, creating in turn a large electric field. The electric field in a 300-micron thick LGAD at 3 different bias values (V = 50 V, 200 V, and 600 V) and that of a PiN diode at V = 600 V are shown in Figure 7, on linear (left side) and logarithmic scales (right side). The electric field in LGAD is therefore clearly divided into two distinct zones: the drift volume with rather low values of the electric field (E ~ 30kV/cm), and a thin multiplication zone located within a depth of a few micrometers with very high field (E ~ 300kV/cm) [12][13]. The implants need to be shaped to allow high bias-voltage operation without breakdown [14]. In the n-in-p LGAD design, electrons drifting toward the $n^{++}$ electrode initiate the multiplication process while in the p-in-n design the multiplication is initiated by the holes drifting toward the $p^{++}$ electrode. Since the multiplication mechanism starts for electrons at a lower value of the electric field than what is necessary for holes multiplication, the n-in-p design offers the best control over the multiplication process. It is in fact possible in the n-in-p design to tune the value of the electric field in such a way that only electrons drive the multiplication process and therefore avoiding operating the device in avalanche mode. Under such conditions the gain is not very sensitive to the exact value of the electric field and the LGAD can be operated very reliably; this condition also minimize the noise coming from the multiplication process, the so called excess noise factor (see section 4.4), enhancing the LGAD performances.

It is important to point out that the internal gain mechanism increases the sensor noise more than the signal, decreasing therefore the signal-to-noise ratio (SNR) of the sensor. However, since the total noise in a silicon detector is dominated by the electronic noise and not by the sensor noise, low values of internal multiplication increase the total SNR. This process is therefore different from the use of an external amplifier, where the amplification works equally on both signal and noise, without a net improvement.

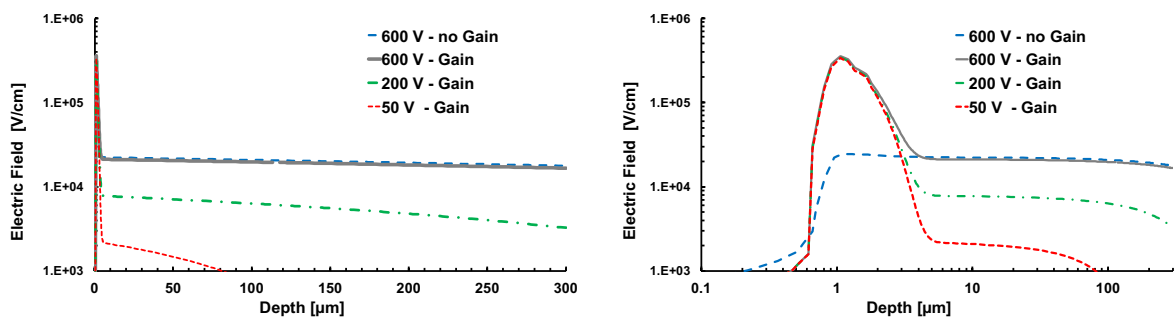

11   Figure 7 The electric field of a 300 µm thick LGAD at different bias voltages compared to a PiN (no gain) Si sensor in linear (left) and logarithmic (right) scale.

### 2.3.3 Why low gain?

High gain silicon devices such as SiPM and APD are designed to have the capability to detect single (SiPM) or a few (APD) photons, respectively, and they need high gain to perform such tasks. However the high value of gain has many drawbacks, namely the increase in sensor noise, the difficulties in sensor segmentation (due to the very high fields), and the high power consumption after irradiation.

Detection of charged particles instead of photons has the advantage of a much larger initial signal, since in one micron 73 electron-hole pairs (MPV) are created by a minimum ionizing particle (MIP), permitting the use of lower gains: the LGAD technology is therefore the solution to the problems caused by high gain. The underlying idea of the UFSD development is to manufacture thin silicon sensors based on the LGAD design that have the lowest gain that is sufficient to perform accurate single particle time measurements.

Leakage current for silicon detectors used in high radiation environments generates shot noise and heat, and it can be the determining factor in the selection of the optimum gain value for such applications, even when cooling the sensor aggressively. As we will see in the following sections, for the low-noise, low-power operation required in a tracker system for a high-energy physics experiment, a gain of $\sim 20$ might be an optimum choice for the operation of LGADs.

### 2.3.4 Why thin sensors?

The current signal generated by a MIP in LGAD has a rather peculiar shape: it has a rise time that is as long as the drift time of an electron traversing the entire sensor thickness, and its maximum current depends uniquely on the value of the gain (see Session 4). For these two facts, assuming a fixed value of gain, the signal steepness depends solely on the sensor thickness: thin sensors have a much faster rising edge, sometimes called "slew-rate", which in turn improves the time resolution. Sensors that are very thin, however, have large values of capacitance and require high gain to generate signals that are large enough to be measured accurately by the read-out electronics: both these facts are detrimental for time resolution. The sensors therefore need to be thin, but not too thin: this delicate balance is explained in the following sections.

Experimental results and simulations indicate that a thickness of $\sim 50$ micron combined with a gain of $\sim 20$ provides optimum performance.

# 3   Detectors optimized for time measurements

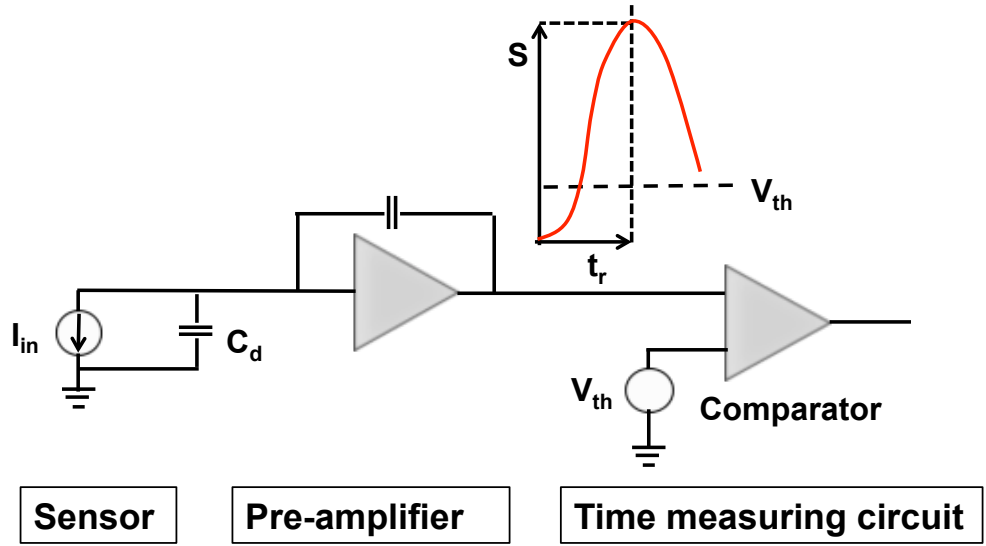

Figure 8 Main components of a time-tagging detector. The time is measured when the signal crosses the threshold.

Accurate time measurements rely on the capability of the read-out electronics to determine the time of passage of a particle using as input the signals generated by the sensor. The most important signal characteristic is to have a constant shape that scales with the amount of energy deposited. As we will see later on in this section, the signal-to-noise ratio (SNR) and the signal slew rate (how fast the signal rises, dV/dt) are also key ingredients, however if the response of the detector varies depending on the impinging particle position, then the time capability is compromised. Figure 8 shows the main components of a time-tagging detector.  For a review of current trends in electronics see for example [15][16]. The sensor, shown as a capacitor with a current source in parallel, is read-out by a preamplifier that shapes the signal. The preamplifier output is then compared to a threshold $V_{th}$ to determine the time of arrival. Not shown in Figure 8 is that the comparator output is then digitized in a time-to-digital converter (TDC).  In the following we will use this simplified model to explore the timing capabilities of various detectors, while we will not consider more complex and space-consuming approaches such as waveform sampling.

In this model, the particle arrival time is defined as the instant $t_0$ when the signal exceeds the threshold: every effect that changes the shape of the signal in the vicinity of the value Vth causes $t_0$ to move either earlier or later and therefore affects the time resolution ($\sigma_t$).

We broadly group the effects influencing the time resolution into 4 categories:

i.  Energy deposition by the particle, determining the amplitude variations ($\sigma_{\text{Time Walk}}$) and the irregularities ($\sigma_{\text{Landau Noise}}$) of the signal

ii.  Signal distortion, due to non-uniform weighting field and varying charge carrier drift velocity ($\sigma_{\text{Distortion}}$)

iii.  Electronics, mostly due to the noise and the amplifier slew rate ($\sigma_{\text{Jitter}}$)

iv.  Digitization, driven by the TDC uncertainties ($\sigma_{\text{TDC}}$)

Following this division, time resolution can be expressed as the sum of several terms, equation (3-1):

$$\sigma_t^2 = \sigma_{TimeWalk}^2 + \sigma_{LandauNoise}^2 + \sigma_{Distortion}^2 + \sigma_{Jitter}^2 + \sigma_{TDC}^2 \qquad (3\text{-}1)$$

Each of these terms will be discussed in details in the following paragraphs using simulation results obtained with the Weightfield2 simulation program (WF2) [17], which will be discussed in detail in Section 4.1, and measurements results in Section 8

## 3.1  Energy deposition: the effect of Landau fluctuations on time walk and signal shape

The ultimate limit to signal uniformity is given by the physics governing the energy deposition by an impinging charged particle in silicon since the total number and the local density of electron-hole pairs created along its path varies on an event-by-event basis. This effect is common to both LGAD and no-gain sensors. These variations not only produce an overall change in signal amplitude, which is at the root of the time walk effect, but also produce irregularities in the current signal (Landau noise). The left part of Figure 9 shows 2 examples of the simulated energy deposition of a MIP in a 200 micron thick no-gain sensor, while the right part displays the associated generated current signals and their components. As the picture shows, the variations are rather large and they can severely degrade the achievable time resolution.

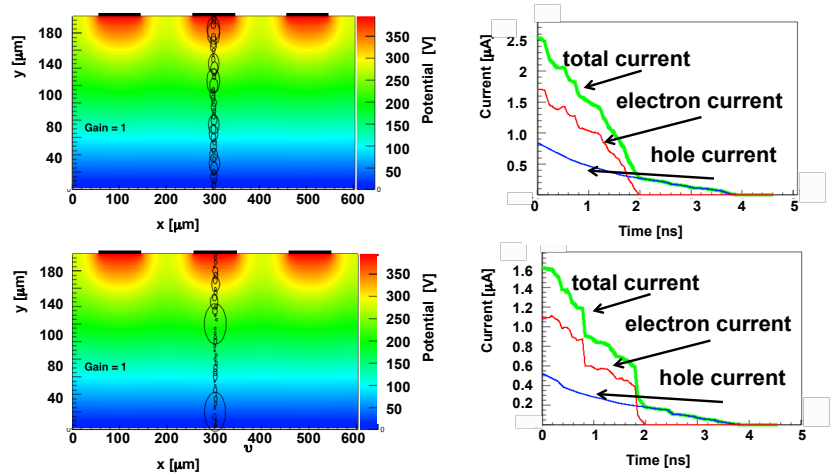

Figure 9 Simulation of the energy deposition by an impinging MIP in a silicon detector and the corresponding current signals. Figure taken with permission from [10].

### 3.1.1 Time walk

The term *Time Walk* indicates the unavoidable effect that larger signals cross a given threshold earlier than smaller ones, Figure 10 left pane. Let's assume for simplicity a linear signal, with amplitude $S$ and rise time $t_r$. This signal crosses the threshold $V_{th}$ with a delay $t_d$, Figure 10 right pane. Using the geometrical relationship $t_d/t_{rise} = V_{th}/S$, the moment when the particle crosses the threshold can be written as: $t_d = \frac{t_{rise}V_{th}}{S}$. Time Walk is then defined as the rms of $t_d$:

$$\sigma_{TimeWalk} = [t_d]_{RMS} = [\frac{V_{th}}{S/t_{rise}}]_{RMS} \propto [\frac{N}{dV/dt}]_{RMS}$$

where we used $S/t_{rise} = dV/dt$, and the fact that the threshold is often expressed as a multiple of the system noise N. Time walk is therefore minimized by systems with low noise and high slew rate. Time walk mitigation techniques in the readout electronic are analyzed in Section 6.4.

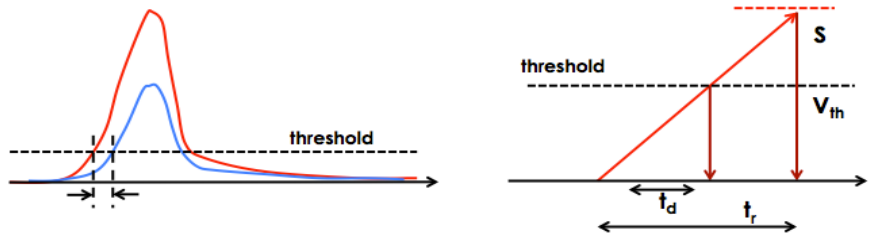

Figure 10 Left side: Signals of different amplitude cross a fix threshold at different times, generating a delay $t_d$ on the on the firing of the discriminator that depends upon the signal amplitude. Right side: a linear signal, with amplitude $S$ and rise time $t_r$ crosses the threshold Vth with a delay $t_d$.

## 3.2   Signal Distortion: non-uniform weighting field and non-saturated drift velocity

In every particle detector, the shape of the induced current signal can be calculated using Ramo – Shockley's theorem [18][19] that states that the current induced by a charge carrier is proportional to its electric charge $q$, the drift velocity $v$ and the weighting field $E_w$, equation (3-2):

$$i(t) = -q \, \vec{v} \cdot \overrightarrow{E_w} \qquad\qquad (3\text{-}2)$$

This equation indicates several key points in the design of sensors for accurate timing:

- the drift velocity needs to be constant throughout the volume of the sensor. The easiest way to obtain uniform drift velocity is to have an electric field high enough everywhere in the active volume of the sensor so that the carriers always move with saturated drift velocity. Figure 11 [20] shows the value of the electron drift velocity as a function of the electric field E for different temperatures. At room temperature, the velocity saturates for a field of about 30 kV/cm, while cooling the sensor has two effects: it lowers the field necessary to reach saturated velocity, and the saturated velocity is higher. Non-uniform drift velocity induces variations in signal shape as a function of the hit position, spoiling the overall time resolution, the effect is shown in Figure 12 a).

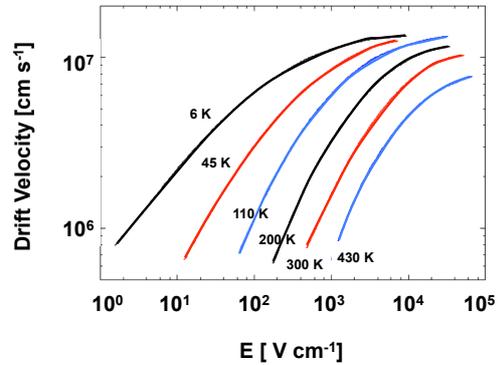

Figure 11 Electron drift velocity as a function of the electric field E for different temperatures. Figure taken with permission from [20].

- the weighting field $E_w$, representing the capacitive coupling of a charge $e$ to the read-out electrode, should not vary along the electrode pitch: if this coupling depended on the impinging particle position along the implant pitch, the signal shape would be different depending on the hit position, spoiling the time resolution. Consider the two cases shown in Figure 12 b): for a wide strip geometry, where the strip width is similar to the strip pitch, the weighting field along the strip pitch is rather constant, while in the case of thin strips, where the strip width is much smaller than the strip pitch, the weighting field is concentrated solely underneath the strip implant.

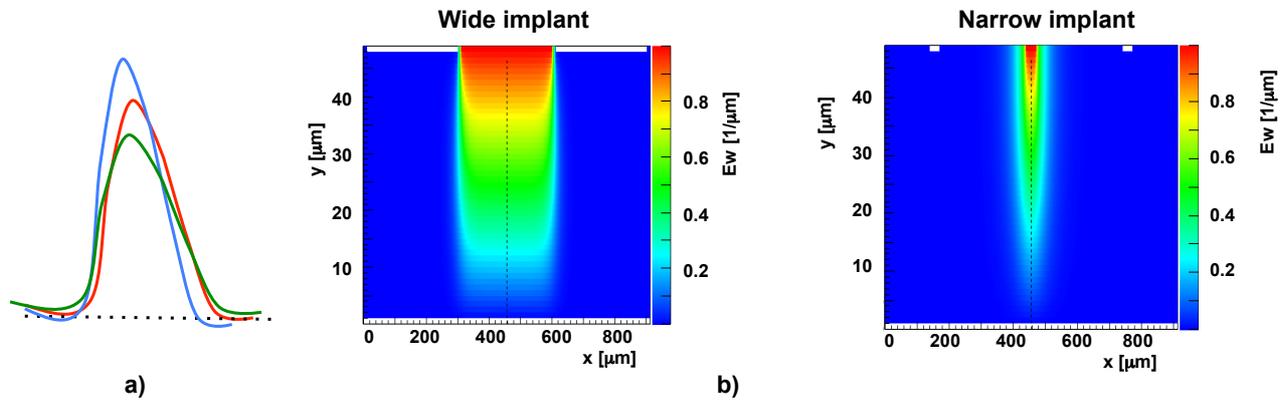

Figure 12 a) Effect of velocity variation on the signal shape b) Weighting field for two configurations: (left) wide implants, (right) narrow implants.

The facts outlined above therefore indicate that to obtain a good time resolution the sensor should have a geometry as close as possible to a parallel plate capacitor, with uniform electric and weighting fields: implants need to have a width very similar to the pitch, and the implant pitch needs to be larger than the sensor thickness: implant width ~ implant pitch >> sensor thickness.

### 3.3    Jitter

The jitter term represents the time uncertainty caused by the early or late firing of the comparator due to the presence of noise on the signal itself or in the electronics. It is directly proportional to the noise $N$ of the system and it is inversely proportional to the slope of the signal around the value of the comparator threshold, Figure 13. Assuming a constant slope, as in Section 3.1.1, we can write $dV/dt = S/t_{rise}$ and therefore:

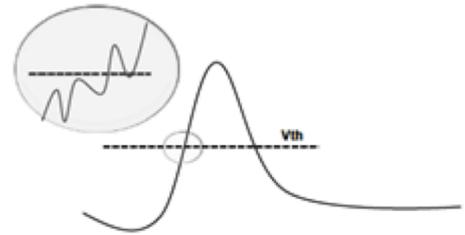

Figure 13 Effect of noise on the crossing of the threshold value Vth

$$\sigma_{Jitter} = \frac{N}{dV/dt} \approx trise/(\frac{S}{N})$$

(3-3)

This apparently simple equation contains the core of the electronic design optimization which is a balance between competing effects: large slew rates require wide bandwidth, which in turn increases the noise while the quest for low noise calls for smaller slew rates. An outline on the characteristics of the front-end amplifiers and the trade-off between noise and slew rate is presented in section 6.1

### 3.4    The "t₀" problem

In systems where the weighting field is not constant over the sensor volume there is an additional source of time uncertainties: before the particle signal can become visible, the charge carriers have to drift from the impact point to the region of high weighting field. This effect is shown schematically in

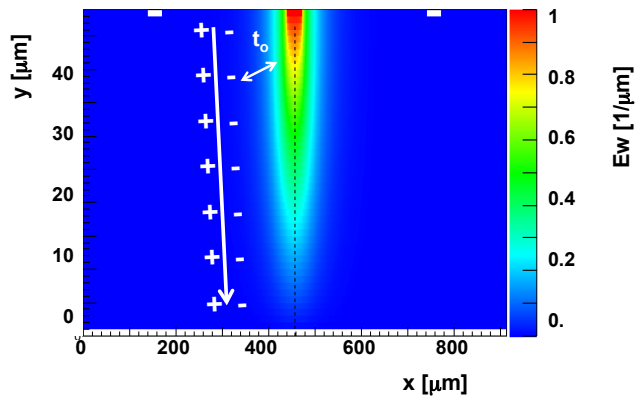

Figure 14 A non-uniform weighting field causes an additional source of time uncertainties due to the drift time from the impact point to the region of high weighting field.

Figure 14. In silicon, electrons with saturated velocity move about 1 μm in 10 ps thus this effect can easily become the dominant source of time uncertainties.

## 3.5 TDC effects on time resolution

The timing information has to be stored for readout. This is typically done in a TDC (time-to-digital converter) where the time of the leading edge of the discriminator signal is digitized and placed in a time bin of width $\Delta T$, given by the TDC least significant bit. This process adds a contribution to the timing uncertainty equal to $\Delta T / \sqrt{12}$, i.e. a bin width of 25 picoseconds will result in a contribution to the overall timing of about 7 picoseconds. The error coming from the TDC is not tied to the sensor characteristics and should be targeted to be a small contributor to the overall resolution. There are other methods to reduce the TDC effect, like employing a template method, fitting the pulse shape or digitizing the entire pulse shape in a sampling digitizer (e.g. SamPic [21]), but most of these methods can't be implemented in a system with a large number of channels.

## 3.6 Summary

Time resolution is the sum of several contributions, equation (3-1):

i.   Jitter and time walk are minimized by detectors with very fast slew rates, with low intrinsic noise and read out by low noise amplifiers.

ii.  The contribution from signal distortions can be minimized by operating the sensor in a regime where the carrier's drift velocity is saturated and by employing a sensor geometry such that the weighting field is uniform. These constraints suggest using "parallel plate" geometries, where the dimensions of the active area are much larger than the sensor thickness.

iii. The time uncertainty caused by non-uniform charge deposition in the active volume, the so-called "Landau Noise", needs to be evaluated in each specific situation, with a detailed simulation of the sensor combined with the read-out circuit.

iv.  The TDC binning represents in most cases a very small effect, and it will be ignored in the rest of this paper.

How to design a sensor that meets the first 3 requirements is the subject of the following sections.

## 4 UFSD Sensor Development

### 4.1 WeightField2: a simulation program for solid state sensors

We have developed a full simulation program, Weightfield2 (WF2) [17], [22], [23], with the specific aim of assessing the timing capability of silicon sensors with internal gain. The program has been validated by comparing its predictions for MIP and alpha particles with both measured signals and TCAD Sentaurus simulations, finding excellent agreement in both cases. Figure 15 shows on the left the comparison WF2-TCAD for the predicted current produced by a MIP in a 300-micron silicon sensors with gain = 14 while on the right the comparison between WF2 and the impulse measured at a beam test with 120 GeV/c pions, using as read-out a charge sensitive amplifier. The left side therefore shows the good agreement in the simulation of the mechanisms involved in the current signal, while the right side shows how the program also correctly simulates the electronic response.

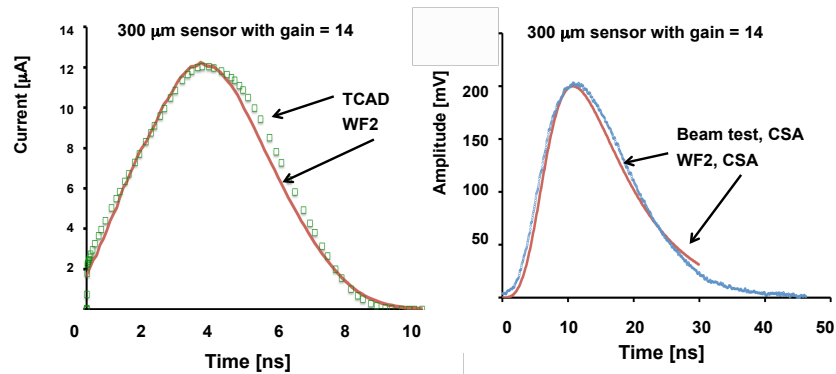

Figure 15 Left: comparison WF2-TCAD for the predicted current produced by a MIP in a 300-micron silicon sensors with gain= 14. Right: comparison between WF2 and the impulse measured at a beam test with 120 GeV/c pions, using as read-out a charge sensitive amplifier.

All the simulation plots and field maps shown in this paper have been obtained with WF2.

### 4.1.1 Description of WeightField2 simulation principles

WF2[3] uses a graphical user interface, shown in Figure 16, for the input of several parameters such as (i) type of incident particle (MIP, α-particle, laser, x-ray), (ii) sensor geometry, (iii) presence and value of internal gain, (iv) doping of silicon sensor and its operating conditions, (v) the values of an external B-field, ambient temperature and thermal diffusion and lastly (vi) the oscilloscope and front-end electronics response. As a debugging tool, WF2 offers also the possibility to display the animated motion of electrons and holes in the sensor.

---

[3] Available for download at http://personalpages.to.infn.it/~cartigli/Weightfield2/Main.html

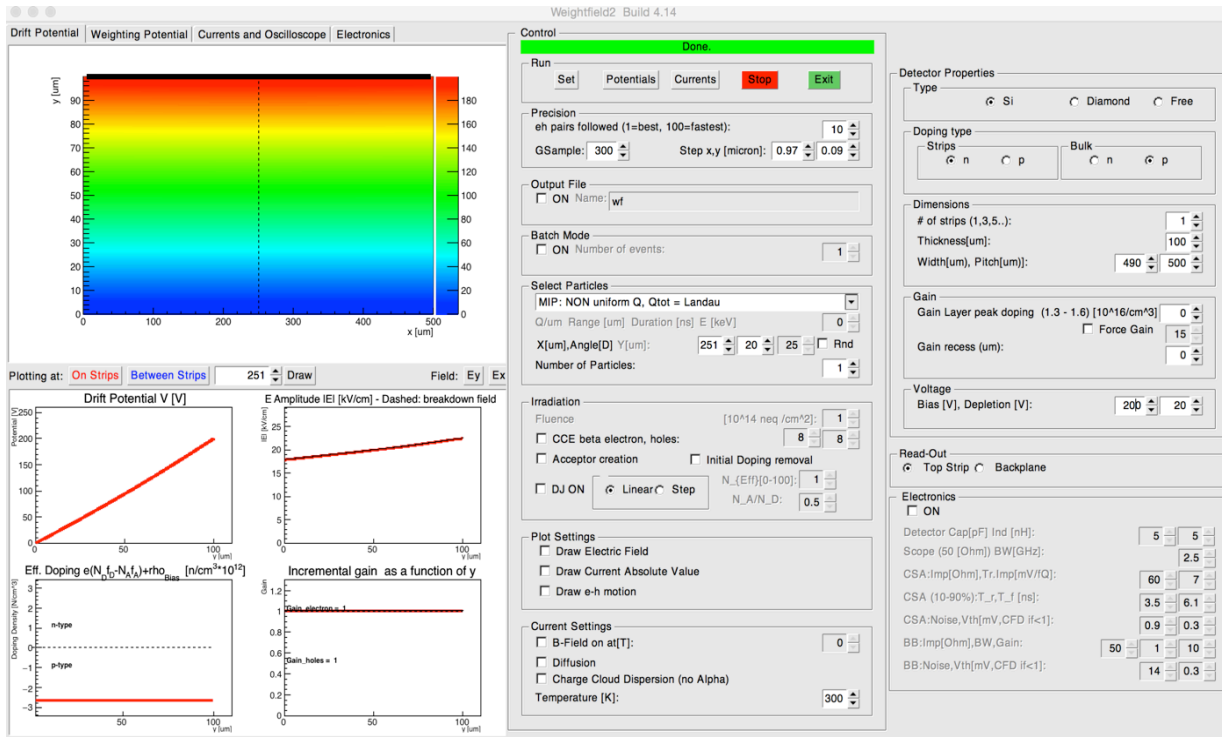

Figure 16 Graphical User Interface of the simulation program Weightfield2.

The electric field and the weighting field are computed by solving Poisson's and Laplace's equations for the related potentials, $\nabla^2 V = -\frac{\rho}{\varepsilon}$ or $\nabla^2 V = 0$, where $\rho$ is the charge density and $\varepsilon$ is the electric permittivity. WF2 performs the calculation iteratively by discretizing the equation on a grid. To obtain a faster calculation, WF2 uses a multi-grid structure, which allows starting the potential calculation on a coarser grid, and then refining it to a grid with halved mesh size at each iteration step. The boundary conditions in the computation of the fields are set by the external applied potential (either $V = V_{bias}$ or $V = 0$ V) at the electrodes, while the other structures in the planes of the electrodes are left floating. In the direction along the x direction, as shown in Figure 16, the boundary conditions are instead periodic.

The program simulates the creation of electron–hole pairs by an ionizing particle by distributing charge carriers along an imaginary trajectory using a library of energy releases computed by the simulation package GEANT4 [24]. With this method, the actual microscopic description of the interactions between a charged particle and the silicon reticule has already been calculated within GEANT4 and therefore WF2 can run faster. The parameterization used in WF2 was obtained using default GEANT4 parameters[4]. The point where a particle hits the detector and the angle formed by the particle with the sensor surface are also selectable in the graphical interface. The drift of the charge carriers generated by an incident particle is followed in time steps, with a precision selected by the user in terms of both the percentage of electron-hole pairs simulated and that of the time unit. For each time step $j$, the induced current $I_{tot}(t_j)$ is derived using Ramo – Shockley's theorem [18][19] by summing over the charge carriers $k$:

$$I_{tot}(t_j) = \sum_{k=1}^{n} I_k(t_j) = -q \sum_{k=1}^{n} \overrightarrow{v_k(t_j, x_k)} \cdot \overrightarrow{E_w}(x_k) \; ; \qquad (4\text{-}1)$$

where $\overrightarrow{v_k(t_j, x_k)}$ is the velocity of the charge carrier $k$ at the time step $t_j$ at the position $x_k$ and $\overrightarrow{E_w}(x_k)$ is the value of the weighting field at the position $x_k$. Table 1 shows the parameterisations used in WF2 for the mobility $\mu_{e,h}(T)$, the $\beta_{e,h}(T)$ coefficients, the drift velocity $v_{e,h}(x)$, and for the saturated drift velocities $v_{e,h,Sat}(x)$.

Table 1 Parameters used in the WF2 simulation.

| | Electrons | Holes |
|---|---|---|
| $\mu(T) \; \left[ {}^{m^2}\!/_{VS} \right]$ | $0.1414 \left( \dfrac{T}{300K} \right)^{-2.5}$ | $0.0470 \left( \dfrac{T}{300K} \right)^{-2.2}$ |
| $\beta(T)$ | $1.09 \left( \dfrac{T}{300K} \right)^{0.66}$ | $1.213 \left( \dfrac{T}{300K} \right)^{0.17}$ |
| $v_{Sat}(T) \; [{}^{m}\!/_{s}]$ | $1.07e5 \left( \dfrac{300K}{T} \right)^{0.87}$ | $8.35e4 \left( \dfrac{300K}{T} \right)^{0.52}$ |
| $v(\mathrm{x}, \mathrm{T}) \; [{}^{m}\!/_{s}]$ | $\dfrac{\mu_e(T) E_d(x)}{{}^{1/\beta_e(T)}\sqrt{1 + (\frac{\mu_e(T)E_d(x)}{v_{e,Sat}(T)})\beta_e(T)}}$ | $\dfrac{\mu_h(T) E_d(x)}{{}^{1/\beta_h(T)}\sqrt{1 + (\frac{\mu_h(T)E_d(x)}{v_{h,Sat}(T)})\beta_h(T)}}$ |

---

[4] For a detailed description of the processes involved in the interaction of a charged impinging particle with a silicon sensor, we refer the interested reader to the GEANT4 manual [24]

where $E_d(x)$ is the drift field at position $x$. The dependence of these quantities upon the absolute temperature T has been taken from the Synopsis Sentaurus manual [25].

WF2 allows the user to insert a gain layer at the p–n junction in order to simulate the behavior of sensors with internal charge multiplication. The gain layer creates a volume in the bulk of the silicon sensor where the electric field is locally high enough ($E \approx 300 kV/cm$) so that the drifting charge carriers will induce a controlled avalanche without a complete electrical breakdown. WF2 also simulates charge multiplication in the bulk due to very high fields at high operating bias.

In high electric fields the moving charges gain sufficient energy that they can initiate ionization, and build up an avalanche over a distance x such that the number of electrons ($N_e$) and holes ($N_h$) grow exponentially (for a detailed review see for example [25] and [26]):

$$N_e(x) = N_e e^{\beta x}; \qquad N_h(x) = N_h e^{\alpha x}, \qquad (4\text{-}2)$$

with

$$\alpha = A_n \, \exp\{-\frac{B_n}{E}\}; \ \ \beta = A_p \, \exp\{-\frac{B_p}{E}\}, \qquad (4\text{-}3)$$

where α and β are the ionization coefficients of holes and electrons respectively (the coefficients α, β represent the inverse mean free path). The terms $A_{n,p}$ are constants to be derived from the experimental fits, while $B_{n,p}$ depends linearly upon the temperature T[5]:

$$B_{n,p}(T) = C_{n,p} + D_{n,p} \, T. \qquad (4\text{-}4)$$

---

[5] The values $A_n = 4.43$ e5 $cm^{-1}$, $A_p = 1.13$ e6 $cm^{-1}$, $C_n = 9.66$ e5 $V \cdot cm^{-1}$, $C_p = 1.17$ e6 $V \cdot cm^{-1}$, $D_n = 4.99$ e2 $V \cdot cm^{-1} \cdot K^{-1}$, and $D_p = 1.09$ e3 $V \cdot cm^{-1} \cdot K^{-1}$ give good fits to the bulk ionization coefficients over the electric field range of $200 - 800 \ kV \cdot cm^{-1}$ and the temperature range of 15–420 K [27].

In WF2 four models of the impact ionisation rate are implemented; two of them, the van Overstraeten [26] and Massey [27] models, are based on the Chynoweth law [26] while the other two, the Bologna [28] and the Okuto – Crowell [29] models propose their own law for $\alpha_{e.h.}$. All models, except for Massey, are also implemented in Synopsis Sentaurus [25], where complete references can be found.

The output of the WF2 simulation consists of the current pulse collected at a selected n++ or p++ electrode, and in addition plots displaying the various components of both current (electrons, holes, and gain carriers) and electric field (Ex and Ey). An optional feature allows simulating the response of both a current and a charge sensitive amplifier, and to visualise the oscilloscope's signal. The program can be used in batch mode, writing the current pulse of each event on a separate file.

## 4.2  Signal formation and its influence on the time resolution

Using Ramo – Shockley's theorem we can calculate the maximum current in a no-gain pad detector of thickness $d$, assuming a saturated drift velocity $v_{sat}$:

$$I_{max} \propto Nq\frac{1}{d}v_{sat} = (n_{e-h}d)q\frac{1}{d}v_{sat} = n_{e-h}qv_{sat} \qquad (4\text{-}5)$$

where we used the fact that $E_w \propto {}^{1}/_{d}$ for a pad geometry, and $N$ is the number of e-h pairs ($N = n_{e-h}\,d$) assuming a uniform charge creation of $n_{e-h}$ e-h pairs per micron. This result shows an interesting feature of silicon sensors: the peak current does not depend on the sensor thickness. Thick sensors have indeed a  larger number ($N$) of initial e-h pairs, however each pair generates a lower initial current (the weighting field is inversely proportional to the sensor thickness d), Figure 17.

This cancellation is such that the peak current in silicon detectors is always the same, $I_{max} \approx 1.5\ \mu A$,

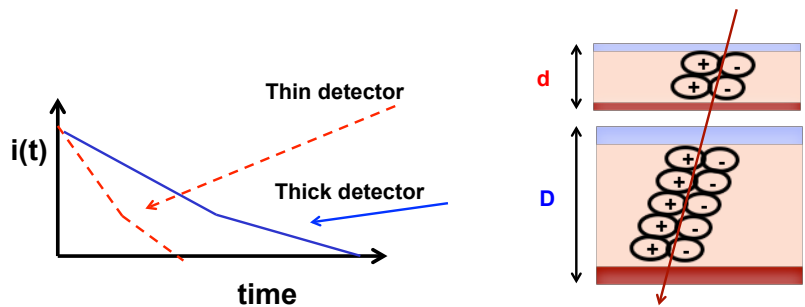

Figure 17 The initial signal amplitude in silicon sensors does not depend on their thickness: thin and thick detectors have the same maximum current, and thick detectors have longer signals.

regardless of the sensor thickness. This fact is at the core of the limited time precision of no-gain silicon detectors: the signal amplitude is limited by the saturation of the velocity and the maximum value of the weighting field, and it cannot be increased using thicker sensors. Current amplifiers, exploiting the fast initial burst of current, give therefore similar time resolutions when used with thin and thick sensors for same noise performance, while charge sensitive amplifiers might benefit from thicker sensors as they integrate the charge over a longer period, however at the price of a longer rise time.

In this paragraph we reached therefore a very important conclusion: the current provided by traditional silicon sensors is not sufficient to obtain very accurate time resolution and it is therefore necessary to introduce internal gain.

## 4.3   The influence of internal charge multiplication on the UFSD output signal

Using WF2 we can simulate the output signal of UFSD sensors as a function of many parameters, such as the gain value, sensor thickness, electrode segmentation, and external electric field. In the UFSD design, as shown in Figure 18, gain is generated only at the moment when an electron enters the gain layer near the n-p junction. This fact is of crucial importance to the understanding of the shape of the UFSD current signal: the current increases as the electrons are drifting towards the anode, so the rise time of UFSD signal is given by the electrons drift time.

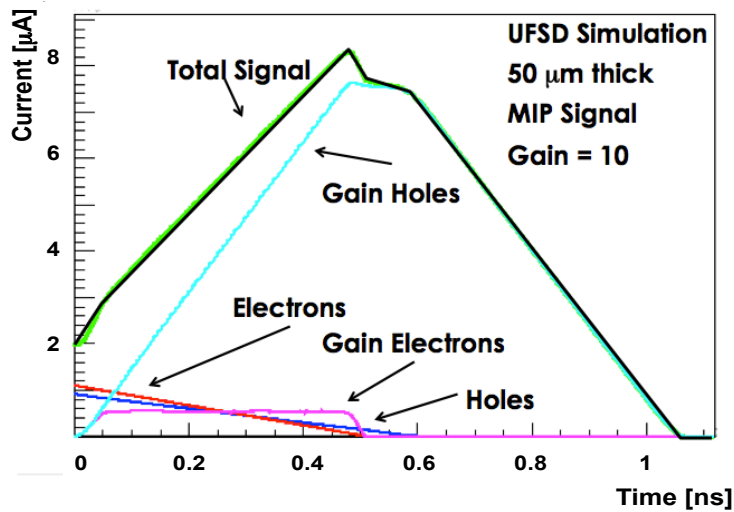

Figure 18 Simulated current signal for a 50-micron thick UFSD. Figure taken with permission from [40].

Figure 18 shows the simulated current, and its components, for a 50-micron thick UFSD. The initial electrons, drifting toward the n$^{++}$ electrode, pass through the gain layer and generate additional e-h pairs. The cathode readily absorbs the gain electrons while the gain holes drift toward the anode and generate a large current.

The production of the additional charges through the multiplication process takes extra time: in this respect 10 particles impinging on a no-gain silicon detector simultaneously will produce a faster signal than one particle with gain 10, since in a no-gain sensor the incident particles create all charge carriers immediately.

## 4.4   Impact ionization mechanism and excess noise factor

For the purpose of this study, a very relevant parameter is the additional noise induced by the multiplication mechanism, the so-called excess noise factor F.  The mechanism at the root of the excess noise factor is explained in Figure 19: each unit charge entering the gain layer generates a number of charges that on average is equal to the gain G, however individually each charge can generates more or less charge. This added noise is such that after multiplication the signal is multiplied by G, while the current noise by $\cdot \sqrt{F}$. Therefore, multiplication improves the SNR only if the dominant noise source is not the current noise. In the literature the excess noise factor is often expressed as a function of gain G and the ratio k, $k = {}^{\alpha}/_{\beta}$:

$$F \sim G^x = Gk + \left(2 - \frac{1}{G}\right)(1 - k), \qquad (4\text{-}6)$$

where x is called the excess noise index, and α (β) is the hole (electron) ionization rate.  Equation (4-6) makes it evident that the key to low noise amplification is low gain coupled to a hole ionization rate as small as possible: the electric field should be such as to cause only electron multiplication. Figure 20 shows the simulated excess noise factor for LGAD detectors according to [30]: the fit to the points yield to a value of k = 0.22 and a value of F in the range 4-6.

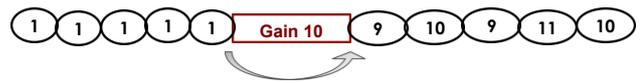

Figure 19: Excess noise factor: each unit charge entering the gain layer from the left generates a number of charges that on average equals the gain G, (10 in this case), but with a spread described by the excess noise factor F.

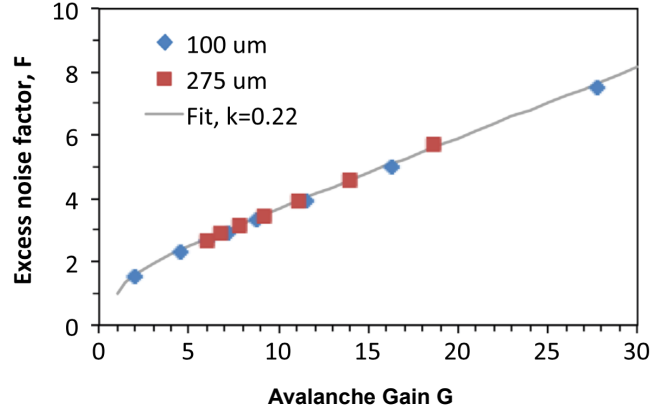

Figure 20: Excess noise factor as a function of gain. Figure taken with permission from [30].

## 4.5 Thin Sensors

The value of the signal current generated by a gain G can be estimated in the following way: (i) in a given time interval dt, the number of electrons entering the gain region is $n_{e-h}\,v_{sat}\,dt$ ($n_{e-h} \sim 73$ e-h pairs per micron); and (ii) these electrons generate $dN_{Gain} \propto n_{e-h}(v_{sat}dt)G$ new e-h pairs. Using again Ramo's theorem, the current induced by these new charges is given by:

$$di_{gain} = dN_{gain}\,qv\left(\frac{1}{d}\right) \propto \frac{G}{d}\,dt, \qquad (4\text{-}7)$$

which leads to the expression:

$$\frac{di_{gain}}{dt} \sim \frac{dV}{dt} \propto \frac{G}{d}. \qquad (4\text{-}8)$$

Equation (4-8) demonstrates a very important feature of UFSD: the increase in signal current due to the gain mechanism is proportional to the ratio of the gain value over the sensor thickness (G/d), therefore thin detectors with high gain provide the best time resolution. Using WF2 we have cross-checked this prediction simulating the slew rate for different sensor thicknesses and gains, Figure 21 left: 300-micron thick sensors with gain 20 have a slew rate a factor of two higher than that of traditional sensors, while a 50-micron thick sensors the difference is more than a factor of 6.

As it was done for non-gain sensor in section 4.2, we can also derive the maximum current for UFSD. In UFSD the number of charge carriers increases until the last electron reaches the anode, and it starts decreasing when the gain holes reach the cathode. Since each electron generates G holes, we can assume that between these two moments the current is:

$$I_{max} \propto N_{max} q \frac{1}{d} v_{sat} = (n_{e-h} dG) q \frac{1}{d} v_{sat} = n_{e-h} G q v_{sat} \qquad (4-9)$$

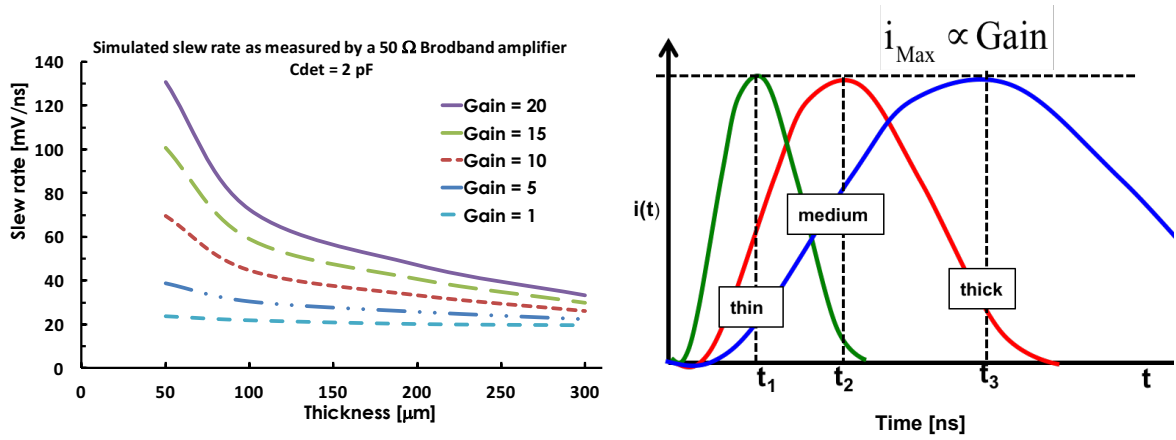

Figure 21 Left: signal slew rate as a function of sensor thickness for 5 different values of gain. Right: current signals from sensors with equal gain and different thicknesses. Figure taken with permission from [40].

where $N_{max}$ is the maximum number of charge carriers, $n_{e-h} \cdot d$ is the number of initial electrons, q the elementary charge, G the gain, 1/d the weighting field and $v_{sat}$ the hole saturation velocity[6].

This expression shows that the maximum current in UFSDs does not depend on the sensor thickness, but only on the gain and the drift velocity. On the other hand, the sensor thickness controls the current rise time, as it is determined by the electrons drift time. Figure 21 shows on the right side both these effects: the gain determines the signal amplitude, the sensor thickness the rise time.

---

[6] For simplicity we assume again a parallel plate geometry, with a uniform charge deposition of 73 e-h pair per micron

## 4.6    Sensors with segmented read-out electrodes

The combination of position and timing information from a single sensor introduces an additional level of complication: position reconstruction favours finely segmented electrodes, leading to very non-uniform electric and weighting fields, while timing measurements require parallel plate geometry, to achieve saturated velocity and minimize signal distortions mentioned in Section 3. Thin sensors offer one possible solution to this conundrum, for example with 50-micron thickness, where the parallel plate geometry can be obtained with rather small segmentation.

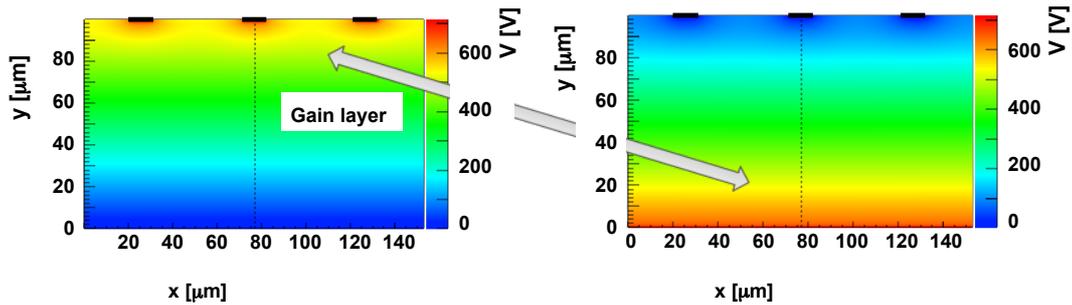

Figure 22 Sketch of the crosscut for two possible configurations of UFSD. Left side: n-in-p configuration, with the gain layer under the segmented electrodes. Right side: p-in-p configuration with the gain layer in the deep side. The secondary y-axis shows the value of the potential. Figure taken with permission from [40].

An interesting option to obtain uniform gain while having segmented electrodes requires using p-in-p sensors [30], where the segmented electrodes and the p-n junction are on opposite sides, Figure 22. In this design the electrode segmentation on the ohmic side of the sensor does not compromise the uniformity of the gain layer, which is at the p-n junction side.

Figure 23 illustrates these two possibilities: on the left side, a thin sensor is read-out via an integrated chip, providing time and position information, while on the right side of the picture, the sensor is read-out by two chips, one for position on one side, and a second one for timing, reading macro pads.

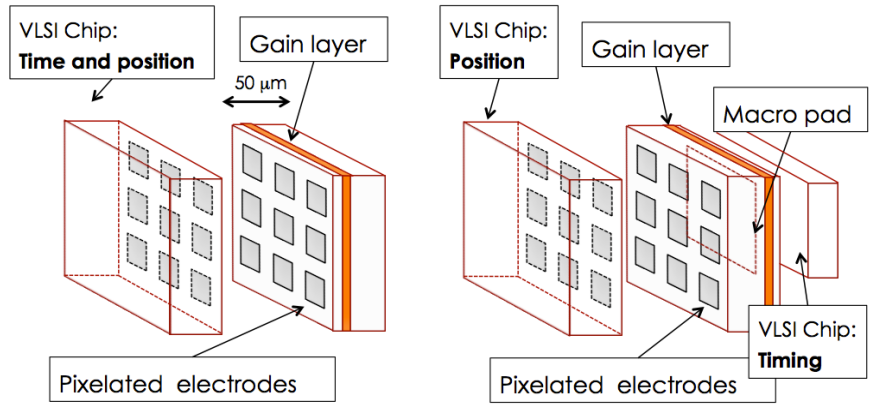

Figure 23 Sketch of a UFSD sensor and associated VLSI electronics. Left side: single read-out chip, right side: split read-out. Figure taken with permission from [40].

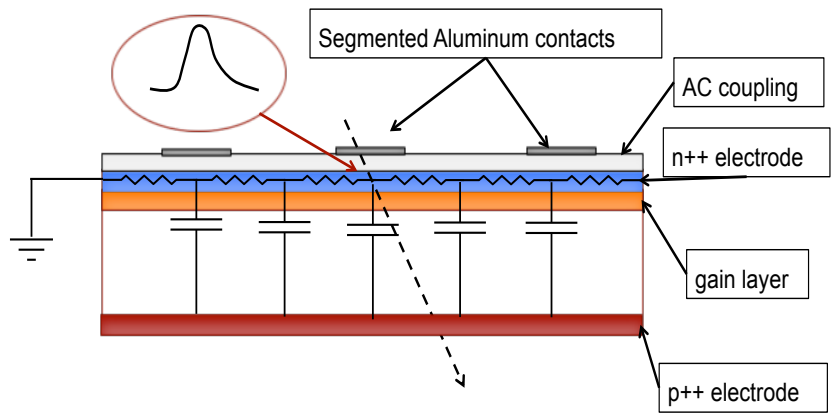

Figure 24 Sketch of segmented read-out via AC coupling

Position resolution, uniform electric field and constant gain can be also obtained using an innovative design, where the segmentation is obtained using AC coupled electrodes. This design is shown in

Figure 24: the continuous anode layer is made slightly resistive so that the signal appears via AC coupling on the aluminium pads, which provide the required segmentation.

## 4.7   Summary

We have developed a fast simulator for silicon sensors with internal gain, WF2. The combination of thin sensor and internal multiplication increases the signal slew rate in such a way that accurate timing becomes possible. Precise timing resolution requires a geometry of the electrodes that is as similar as possible to that of a parallel plate capacitor, with the sensor thickness much smaller than the pad size. The segmentation of the electrodes, necessary to provide concurrent accurate time and position resolutions, should be designed in such a way not to spoil the uniformity of both drift and weighting fields.

# 5    Radiation hardness of UFSD

Many applications of UFSD will be conducted in an environment characterized by radiation. Since UFSD are silicon sensors, we can expect radiation damage effects, depending on the type and level of the radiation. All detectors at the LHC are using silicon sensors on a large scale and planning to use them on an even larger scale for the upgrade to the HL-LHC, and thus these effects are being thoroughly investigated for sensors with no gain by the CERN collaborations. The RD50[7] collaboration is researching radiation damage in UFSD, and finds in the bulk the same effects as in no-gain sensors, somewhat modified to account for the effect of the gain. In addition, for UFSD the effect of radiation on the gain mechanism has to be explored. Proposed applications at the HL-LHC will require functioning of the UFSD after fluences of $5e15\ n_{eq}/cm^2$ and ionizing doses of about 150 Mrad [31].

## 5.1    Radiation effects in silicon sensors with no gain

The properties of silicon sensors are determined by the existence of energy levels within the band gap. Aside from surface effects due to the ionizing part of the radiation, which tend to saturate at the Mrad level, the main damage is to the bulk of the sensors caused by the non-ionizing energy loss (NIEL) which introduces either neutral or charged defects within the band gap. Figure 25 [32] shows the effects graphically: neutral mid-gap effects generate leakage currents, shallow defects trap charges and reduce the collected charges, and shallow and deep charged defects change the doping concentration $N_{eff}$.

As discussed in more detail below, radiation can change the balance of donors and acceptors through the interaction of the dopants with damages to the lattice, i.e. interstitials and vacancies.

The resulting changes in detector properties are proportional to the particle fluence and depend critically on the thickness of the sensor. A thorough review of radiation effects in n-type silicon including the properties of silicon defect engineered with oxygen and carbon was published by the Rose collaboration in 2001 [33]. The advantages of p-type silicon were demonstrated with respect to reduced adverse annealing and reduced trapping [34]. A recent review [32] relates the density of defects in the band gap to macroscopic radiation effects.

---

[7] http://rd50.web.cern.ch/rd50/

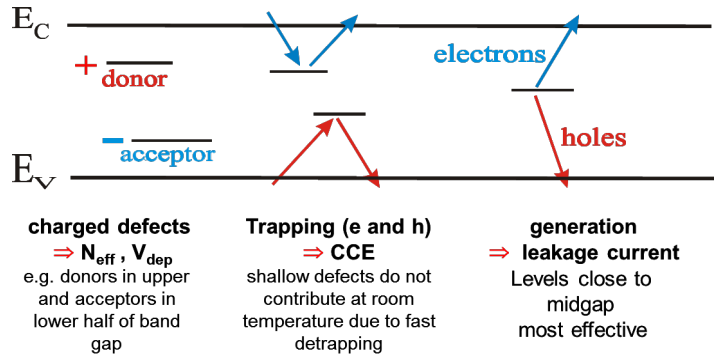

Figure 25 Effects of defects on the properties of silicon sensors

### 5.1.1 Leakage current increase

The observed increase in leakage current is proportional to the number of defects introduced in the lattice by a fluence Φ and to the detector volume V:

$$\Delta i = \alpha \cdot V \cdot \Phi \qquad (5\text{-}1)$$

with α = 2.5e-17A/cm for protons and α = 4.0e-17 A/cm for neutrons, respectively.

This effect depends exponentially on the temperature since the nature of the defect-related recombination mechanisms inducing leakage current is exponentially T-sensitive:

$$i(T) = i_o T^2 e^{\frac{1.2\,eV}{2kT}} \qquad (5\text{-}2)$$

and can thus be mitigated by lowering the operating temperature. Lowering the temperature by $7^0$ C reduces the leakage current approximated by a factor of two.

### 5.1.2 Charge collection efficiency

The probability of trapping during the drift of the charge carriers in the silicon bulk increases with fluence. It is determined by the trapping time, which for a fluence of Φ = 1e15 $n_{eq}$/cm² and the maximum drift velocity corresponds to a distance of about 50μm. Thus at large fluences only sensors of up to that thickness will collect a majority of the charges created.

The trapping-induced decrease of signal has been modeled following an exponential fashion [31][35][36]:

$$I(t) = I_0 e^{-\frac{t}{\tau_{eff}}} \tag{5-3}$$

where $\tau_{eff}$ refers to the effective trapping time, which is inversely proportional to the fluence $\phi$: $\frac{1}{\tau_{eff}} = \beta\phi$.

The value of the parameter $\beta$ can be experimentally determined, and is usually found to differ between electrons and holes. This effect is implemented in WF2: the program uses equation (5-3) to determine at each time step in the propagation of the charge carriers if trapping has happened. De-trapping is not yet included in WF2.

The effect of trapping can be seen in Figure 26 [37] where the collected charge is shown to decrease as a function of fluence for 300μm thick sensors. It should be pointed out that the change in collected charge is much less for thin detectors.

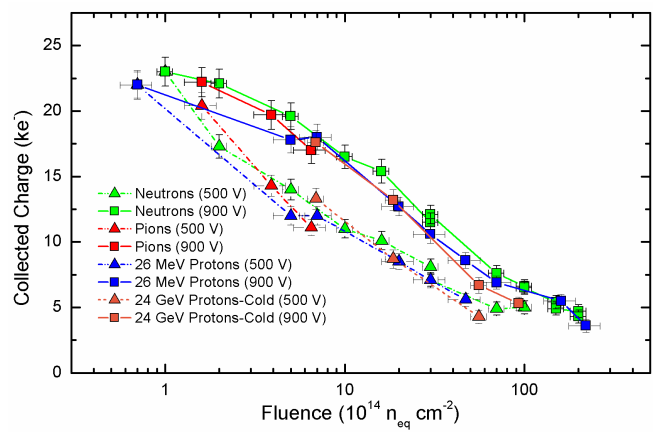

Figure 26 Collected charge for 300 μm thick n-on-p sensors as a function of fluence. Figure taken with permission from [37]

### 5.1.3 Changes in doping concentration

Acceptor creation by deep traps and initial acceptor (Boron) removal [38] in silicon detectors can be parameterized according to the following expression:

$$N_A(\phi) = \ g_{eff}\,\phi + \ N_A(0)\,e^{-c(N_A(0))\phi} \qquad\qquad (5\text{-}4)$$

where $\phi$ is the irradiation fluence [particles/cm$^2$] and $N_A(0)$ is the initial acceptor density [n/cm$^3$]. The first term of equation (5-4) accounts for acceptor creation by deep traps ($g_{eff}$= 0.02 cm$^{-1}$) while the second term for the initial acceptor removal, where the factor c($N_A(0)$) depends on the initial acceptor concentration $N_A$ (0). Figure 27 shows the value of c($N_A(0)$) as a function of initial Boron concentration. The experimental points shown on Figure 27 have been taken from [38] and from presentations at TREDI 2017[8]. WF2 uses the parameterizations shown on the plot to account for this effect.

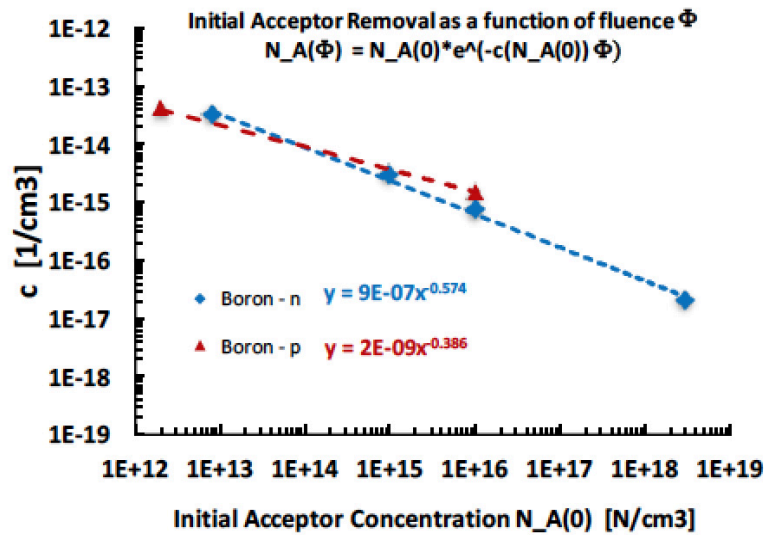

Figure 27 Boron removal coefficient as a function of the initial Boron concentration for proton and neutron irradiation. The coefficient is smaller for larger initial concentrations and it is larger for charged particles.

---

The change in doping concentration has also direct consequences on the full depletion voltage $V_{FD}$, defined as

$$V_{FD} = \frac{qN}{2\varepsilon}d^2, \qquad\qquad (5\text{-}5)$$

with q the electron charge and d the sensor thickness, since the doping concentration N is a function of the fluence. The value of $V_{FD}$ increases with fluence however, due to breakdown effects, the sensor might not sustain the full depletion voltage, with detrimental effects on the sensor performance. There is however a beneficial effect of radiation damage on the breakdown voltage: irradiated sensors tend to have increased breakdown voltages and in thin sensors a large over-depletion can be maintained even after large values of fluence.

## 5.2   Radiation effects specifics to LGAD

For LGADs the radiation effects in the previous section still hold true, including the advantages of thin sensors, with two notable additions: (i) the increased leakage current is multiplied by the gain value, and (ii) the changes in doping profile affect the gain value.

### 5.2.1   Effect of increased leakage on power consumption and shot noise

In LGADs the leakage current generated in the bulk is multiplied by the gain factor G before being collected at the electrodes:

$$i_{LGAD} = G \cdot i_{no-gain}. \qquad\qquad (5\text{-}6)$$

This has immediate consequences for the operation of the UFSD, increasing the required power P = $i_{LGAD}*V_{BD}$ for voltages at break-down by about a factor of the gain G:

$$P_{LGAD} = G \cdot P_{no-gain}. \qquad\qquad (5\text{-}7)$$

The power can be reduced with thinner sensors, since both the leakage current and the operating voltage in thin sensors are lower than in thick ones, and by cooling as mentioned before, but since cooling of the sensors tends to be fairly complicated and expensive, the power consideration might put a limit on the acceptable value of the gain.

Another consequence of the leakage current growth with radiation is the increase in shot noise: shot noise arises when charge carriers cross a potential barrier, as it happens in silicon sensors, and is due to

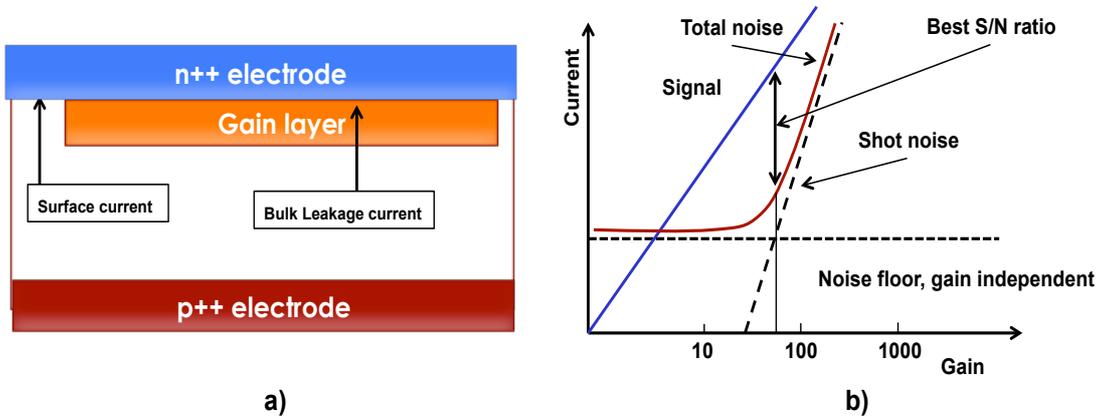

**a)**                                            **b)**

Figure 28 a) Sketch of the shot noise mechanism in sensors with internal gain: bulk current is multiplied by the gain, while surface current is not. b) For increasing gain, shot noise increases faster than the signal.

the finite fixed charge of each electron. As the leakage current increases, so does the shot noise. In sensors such as UFSD this effect is enhanced by the gain and for this reason shot noise can be the dominant source of noise for detectors with gain. As shown in Figure 28 a), the sensor leakage current is the sum of two components: (i) surface current, that does not go through the multiplication layer, and (ii) bulk current, that is multiplied by the gain mechanism.

As discussed in Section 4.4 and shown in Figure 19 when carriers undergo multiplication there is an additional mechanism that enhances shot noise, the so-called excess noise factor (ENF). This effect is in addition to the increase in noise due to the gain value that simply multiplies the leakage current. ENF causes a very peculiar effect: in devices with gain, the signal is multiplied by the gain value while the noise by the gain and by the ENF values: as the gain increases, the SNR becomes smaller since shot noise increases faster than the signal. The consequences are shown in Figure 28 b): the effect of gain is beneficial only up to the point when shot noise becomes the dominant source of noise while after that point the effect of the gain is to decrease the detector performances in terms of the SNR. The shot noise current density is given by:

$$i_{Shot}^2 = 2qI_{Det} = 2q[I_{surface} + (I_{Bulk} + I_{Signal})G^2G^x],$$     (5-8)

where q is the electron charge and $G^x$ is the excess noise factor expressed as a power of the gain value.

Shot noise is normally much smaller than the electronic noise floor for un-irradiated sensors, but it can become the dominant source of noise for irradiated detectors. As an example, Figure 29 shows the

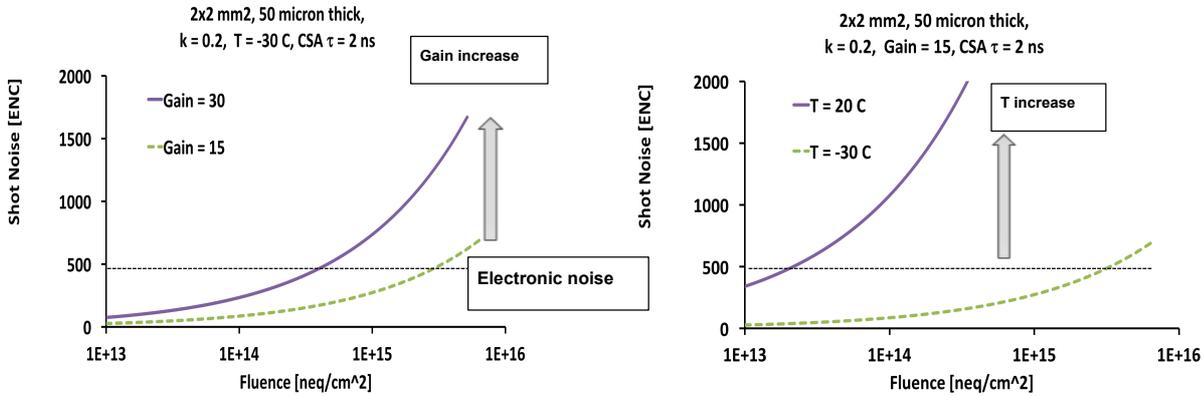

Figure 29 a) Shot noise increase as a function of fluence for two different gain values. b) Shot noise increase as a function of fluence for two different temperature values value of shot noise for a 2x2 mm$^2$ 50-micron thick silicon sensor, assuming a read-out based on a CSA with a 2-ns long integration time. In the plots the electronic noise is assumed to be ENC = 500 e$^-$ and k = 0.2, as calculated in Figure 20. Figure 29 a) illustrates the dramatic effect of gain on shot noise, while Figure 29 b) the effect of temperature, demonstrating that shot noise can become the most important source of noise for irradiated sensors with gain, and it strongly suggests that low gain, short shaping time (details in Section 6.2) and low temperature are necessary for low noise operation.

### 5.2.2 Changes to the doping concentrations

The changes in doping concentration under irradiation in LGAD sensors have been described in Section 5.1.3. The effect of acceptor creation and initial acceptor removal described by equation (5-4) can be applied to the multiplication layer as well as to the bulk. This dependence is shown schematically in Figure 30: the initial Boron doping is removed as the fluence increases and in the meantime new acceptor states are created. The initial Boron removal rate is higher for lower initial concentrations.

At sufficiently high fluence values all curves of the different initial doping concentration converge on the same straight line of the high resistivity PiN diodes, indicating a complete disappearance of the initial donor density.

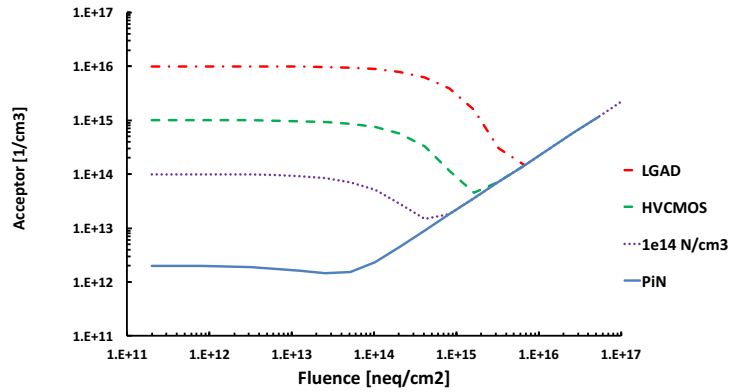

Figure 30 Acceptor concentration as a function of fluence for different initial doping concentrations

Invoking equation (5-4), the change of depletion voltage for LGAD can be calculated; this is shown in Figure 31 for an LGAD consisting of a thin multiplication layer of doping concentration N = 2e16/cm$^3$ and a 45-micron thick high-resistivity bulk with doping N = 2e12/cm$^3$. The full depletion voltage is the sum of the two terms.

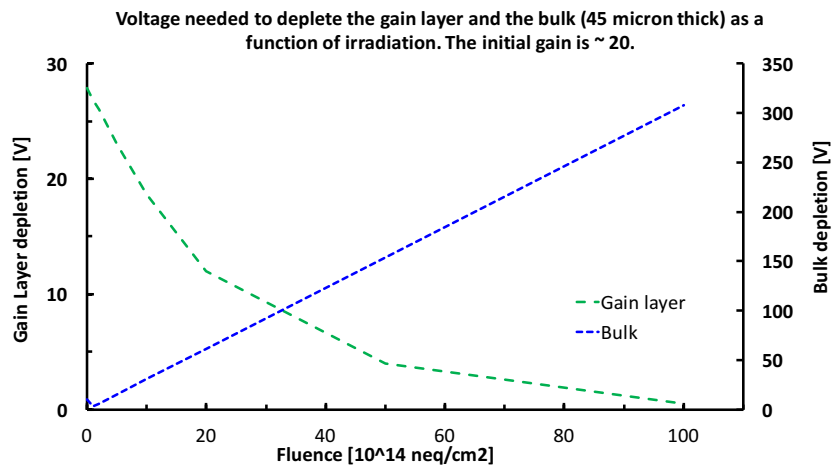

Figure 31 Depletion voltage as a function of fluence for the gain layer and the bulk for a typical 45μm thick UFSD

The root cause of initial acceptor removal is still under investigation [38], however simulations and experimental evidence seem to exclude known radiation effects like "Space Charge Inversion" or "Double-Junction" as the cause for the apparent reduction in the doping concentration.

Interestingly, physical removal of boron atoms can be excluded from kinetic considerations and direct SIMS measurements. A possible explanation of this effect invokes the existence of interstitial defects (I) in the irradiated silicon, well established by RD50 research, with a fluence dependent concentration that form a B-I complex with boron atoms which is electrically inert. These complexes make the Boron atom electrically inactive.

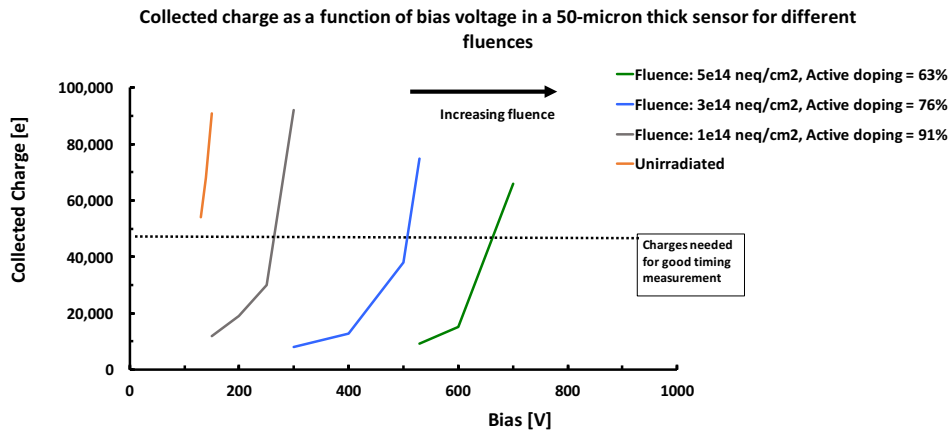

Figure 32 Scenario of gain restoration by increasing the bias voltage as a function of fluence. The lines represent gain vs. bias voltage curves at increasing fluence levels.

This explanation has motivated two avenues of research within RD50 to mitigate the reduction of gain in irradiated UFSD: (i) reduce the concentration of interstitials available for capturing B atoms by using carbon enriched wafers where the interstitials get filled with C instead of with B, and (ii) reduce the formation of the acceptor-interstitial by replacing Boron with Gallium [39].

Another approach, which proved to be successful, is to balance the gain generated by the multiplication layer and by bulk biasing, respectively. This is based on the fact that the gain depends on the electric field and that the field is supplied both by the gain layer and by the field in the bulk. Before radiation, almost exclusively the UFSD gain layer provides the field while with increased fluence, when the increased acceptor removal changes the effectiveness of the gain layer, the higher bias compensates this loss. This technique is possible since after irradiation UFSD can be operated at higher bias values compared to pre-rad operation and an electric field strength supporting charge multiplication are possible in the bulk. This is schematically seen in Figure 32.

### 5.2.3 Changes in output signal shape due to trapping and initial acceptor removal

The simulated combined effect of charge trapping and initial acceptor removal on the UFSD output pulse is shown in Figure 33. The first plot on the left shows the induced current pulse for an un-irradiated sensor, 50-micron thick with gain ~ 10, the middle plot the current after a fluence $\Phi = 6e14$ $n_{eq}/cm^2$ and the right plot after $\Phi = 2e15$ $n_{eq}/cm^2$.

Trapping decreases the current increasingly at longer drift time, while the changes of the location where multiplication happens, from the gain layer to the bulk, affects the shape of the induced current signal since the contribution from gain electrons starts to be relevant.

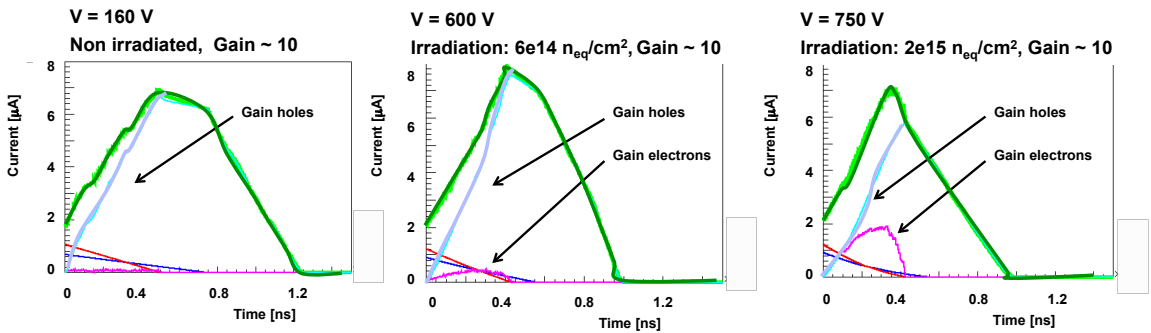

Figure 33 Simulated combined effect of charge trapping and initial acceptor removal on the UFSD output pulse.

Compared to thick no-gain sensors shown in Figure 26, in UFSD the overall changes with radiation are fairly mild, indicating the possibility of performing accurate timing even after high values of fluence. Notably, the overall signal length decreases slightly due to trapping, and the rise time becomes shorter since the current plateau due to holes current disappears.

## 5.3   Summary

Radiation damage causes three main effects in UFSD: (i) decrease of charge collection efficiency, (ii) increase of leakage current, and (iii) changes in doping concentration. Albeit these aspects are common to all silicon sensors, the presence of the gain layer makes UFSD particularly sensitive to the increase of leakage current and the changes in doping concentration. Shot noise can be kept under control using pads insisting to small volumes, in addition to cooling the sensors, and changes in doping concentrations can be compensated using higher post-irradiation values of the bias voltage.

# 6 UFSD Read-out Electronics

## 6.1 Choice of preamplifier architecture

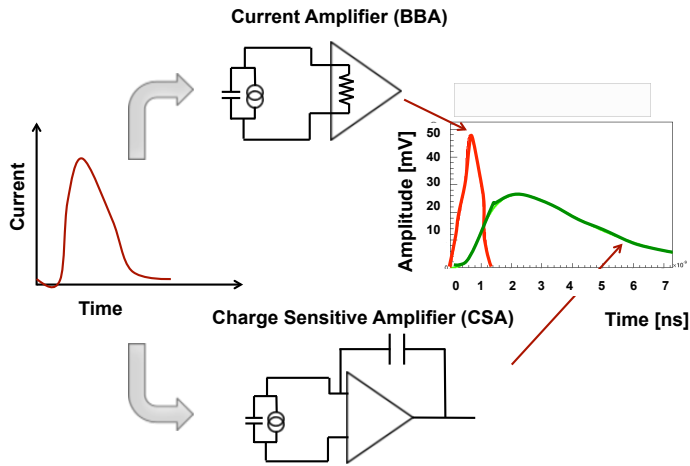

Figure 34 Sketch describing how BBA and CSA amplifiers respond to the same input signal

The sensor readout preamplifier electronics can be broadly broken down into two simplified categories: (i) current mode amplifiers, (usually called broad-band amplifiers, BBA, as they offer a large bandwidth) (ii) charge sensitive amplifiers (CSA) [40]. With BBA the signals are amplified without strong additional shaping while with CSA the signals are integrated and shaped. There are several issues that need to be considered when using either approach: BBAs are much faster, and they are able to take full advantage of the very fast signal slew rate but they have a higher noise, while CSA are somewhat slower but the integration they perform makes the output signal more immune to noise and Landau fluctuations. Figure 34 shows how BBA and CSA threat the same input signal: BBA output is faster and steeper while CSA output is slower and less noisy.

### 6.1.1 Broad-Band Amplifiers BBA

Broad-band (also called current-mode) Amplifiers translate the current signal from the sensor into a voltage signal with some gain, so that $V(t) \sim i(t)$, and they require a wide bandwidth to approximately follow the time structure of the current sourced by the sensor. The slew-rate is then given by dV/dt ~ di(t)/dt, which maximizes the advantage of using a thin sensor for which the derivative of the current pulse is large. For this amplifier the jitter is minimized by the large dV/dt term while the noise term tends to be quite large due to the wide bandwidth. The minimization of the jitter requires the use of high current, however there are often limitations on the power available.

An RC circuit with a current source in parallel to the capacitor can approximate a silicon detector read-out by a BBA; in this approximation, the circuit has a time constant $\tau = R_{in}C_{det}$ where $C_{det}$ is the detector capacitance and $R_{in}$ the read-out input impedance. This means that it takes a time $t_{rise} \sim 2.2\ \tau$ for the current to fully develop the equivalent voltage across $R_{in}$: in order to fully exploit the very high slew rate offered by UFSD, $\tau$ has to be shorter or, at most, of the same order of the signal rise time, $t_{rise}$. The time constant $R_{in}C_{det}$ acts therefore as a reduction of the bandwidth with increasing detector capacitance.

This constraint strongly links sensor and electronics designs, as the electronics should be designed such that it does not slow down very fast input signals. Generally, the ideal BBA has therefore low input impedance and high output impedance. An attractive technology for the BBA preamplifier is silicon germanium bipolar technology that can achieve very high bandwidth and low input impedance.

### 6.1.2 Charge Sensitive Amplifiers CSA

Charge sensitive amplifiers employ a degree of integration of the signal and generate a voltage output signal proportional to the charge collected on the feedback capacitor ($C_f$) of the preamplifier stage; the integration time is controlled by the $R_fC_f$ time constant of the feedback circuit. CSA minimizes the jitter by virtue of a small noise term, as the dV/dt is typically less steep than for the current amplifier. As in an integrator the term dV/dt $\sim$ dQ(t)/dt = i(t), the slew rate is maximized in CSA when the current is at its maximum. For this design the advantage of using thin sensors comes primarily from the fact that the current pulse is short allowing for a short integration and not from the steepness of the signal.

The output shape signal of a CSA is governed by two time constants:
- The rise time: $t_{rise} \sim (C_{Det} + C_{Load})/g_m$ where $g_m$ is the input stage transconductance and $C_{Load}$ is the capacitance value of the circuit loading the preamplifier output
- The fall (discharge) time: $t_{fall} \sim R_fC_f$, controlled by the feedback components.

The fall time, $t_{fall}$, should be longer than the rise time, $t_{fall} \gg t_{rise}$, otherwise the charge will discharge too quickly from $C_f$ and, the peak voltage $V_{out}^{peak}$ instead of reaching the full amplitude $\frac{Q_f}{C_f}$, will only reach a smaller value given by the expression:

$$V_{out}^{peak} = \frac{Q_f}{C_f}(\frac{t_f}{t_r})^{[t_r/(t_r+t_f)]} \qquad (6\text{-}1)$$

This effect is called ballistic deficit [15].

The value of the detector capacitance has a strong impact on CSA performances as (i) it increases the noise, (ii) it slows down the amplifier rise time, $t_{rise} \sim (C_{Det} + C_{Load})/g_m$, and (iii) it decreases the signal amplitude since the fraction of signal charge $Q_s$ stored on the feedback capacitor $C_f$ (and therefore actively contributing to the signal amplitude) depends on the detector capacitance according to the following relationship:

$$\frac{Q_f}{Q_s} = \frac{Q_f}{Q_{Det} + Q_f} = \frac{(1 + A_o)C_f}{C_{Det} + (1 + A_o)C_f} \qquad (6\text{-}2)$$

where $A_o$ is the input transistor open loop gain.

An attractive choice for the CSA amplifier is a small feature size CMOS technology, able to minimize the noise term entering into the jitter.

A comparison of BBA and CSA is shown in Table 2.

Table 2 Comparison of BBA and CSA properties. The symbols used are defined in the text.

| | Slew rate | Rise time | Fall time | Ballistic deficit | Available charge |
|---|---|---|---|---|---|
| **BBA** | $di/dt$ | $R_{in}C_{Det}$ | $R_{in}C_{Det}$ | 1 | $Q_{Tot}$ |
| **CSA** | $i(t)$ | $(C_{Load}+C_{Det})/g_m$ | $R_f C_f$ | $\frac{Q_f}{C_f}(\frac{t_f}{t_r})^{[t_r/(t_r+t_f)]}$ | $(1+A_o)C_f/C_{Det}+(1+A_o)C_f$ |

## 6.2 Noise considerations

The left side of Figure 35 shows the basic blocks of a silicon sensor read-out by a generic amplifier.

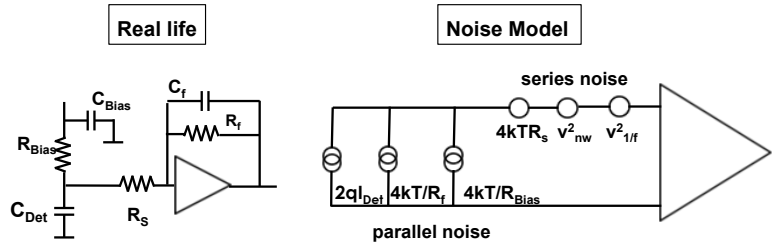

Figure 35 Noise model for a sensor read-out by a generic amplifier.

The circuit has several noise sources such as sensor leakage current, resistor thermal noise, amplifier white and flicker noise; each noise source can be modelled by a function that describes the power generated by the noise source as a function of frequency, the so called spectral power density referred to the input. Capacitors, resistors and the sensor leakage current have spectral density functions that do not depend on the specific circuit, while the amplifier has its own specific spectral density functions. The right part of Figure 35 shows the power density functions of various components: following the standard convention, the noise sources are grouped into parallel noise sources (detector leakage current, feedback and biasing resistor thermal noise), modelled as a current source in parallel with the amplifier input, series noise sources (series resistors thermal noise), modelled as voltage sources in series with the amplifier input, and two voltages sources to model the amplifier white and flicker noise (see for example the discussion presented in [15]). Finally, the actual contribution to the total noise from each source is obtained by calculating the convolution of the power density function with the amplifier transfer function and therefore depends on the amplifier specific transfer function (it is worth noting that the amplifier transfer function used in the noise calculation is specific to the noise source, and in general does not coincide with the signal transfer function). Each noise contribution is quoted then as the number of electrons needed at the input to produce the same voltage signal as the noise does, the so-called equivalent noise charge (ENC). This is directly applicable to the CSA that measures the charge collected from the sensor. In addition the CSA determines the time constant internal to the circuit, which features in analyses of the noise below through an amplifier peaking time $\tau_p$.

In a very general way, we can express the total equivalent noise charge for a CR-RC shaper [15] as the sum of three contributions: series noise, parallel noise and flicker 1/f noise:

$$ENC = \sqrt{ENC_s^2 + ENC_p^2 + ENC_f^2}$$ (6-3)

where:

- $ENC_s \propto \sqrt{\frac{1}{\tau_p}} \; (C_{Det} + C_{in})$
- $ENC_p \propto \sqrt{\tau_p}$,
- $ENC_f \propto (C_{Det} + C_{in})$
- 

and $\tau_p$ is the amplifier peaking time, $C_{in}$ groups together all capacitors in parallel with the detector. For our purposes, the key points of this expression are:

- The effect of series noise ($ENC_s$ and $ENC_f$) is directly proportional to the input capacitance; therefore sensors presenting a small capacitance to the front-end amplifier are preferable, while parallel noise does not depend on the input capacitance. The flicker noise is technology dependent but very small for the fast circuits used for timing.
- The term $ENC_p$ is the sum of the thermal contributions from all the resistors in parallel with the input amplifiers plus the contribution from the sensor leakage current. For this last contribution, the so-called shot noise, the spectral density is given by $i_n^2 = 2qI_l(G^{2+x})$, where $I_l$ is the detector bulk leakage current, $(G^{2+x})$ the factor due to the gain mechanism, and q the electron charge. For CSA, integrating the power density leads to the expression: $ENC_p = \sqrt{\frac{I_l\tau_f}{2q}} \, G^{1+x/2}$, where $\tau_f$ is the feedback time constant, $\tau_f = R_f C_f$. For BBA instead the expression is $ENC_p = \sqrt{\frac{I_l\tau_p}{q}} G^{1+x/2}$ where $\tau_p$ is the amplifier peaking time.

In the operation of UFSD, given the very fast shaping time, the series noise dominates at low values of leakage current while the parallel noise might become dominant after high values of fluences.

## 6.3 Signal collection time and amplifier shaping time

As we have seen in section 3.3, the jitter contribution depends on the noise and on the preamplifier rise time. In traditional (no-gain) silicon sensors, the rise time of the signal is extremely fast, it is the time of passage of the particle in the sensor, and therefore the preamplifier uniquely determines the rise time. UFSD, on the contrary, generates a signal with rise time that is determined by the electron drift time and sensor thickness. Under this condition, a BBA amplifier with a shaping time much faster than the signal rise time, $t_{shaping} < t_{rise}$ will not benefit from the added slew rate, and it will be penalized by the increased noise. On the other hand an amplifier that is too slow, $t_{shaping} > t_{rise}$, will have a significantly slower pulse rise-time coming out of the amplifier. This will result in significantly reduced performance since the peak signal height is reduced in this case and both the amplified pulse rise-time and the signal height enter into the slew rate determination.

A practical choice for the amplifier input impedance $R_i$ is that the product $R_i C_{det}$ be equal to the intrinsic signal rise time, which makes then $R_i$ dependent on the detector capacitance.

## 6.4 Time-walk correction circuits and shape irregularity mitigation

As it was explained in section 3, the variability of the energy deposition by impinging particles creates two distinct effects: amplitude fluctuations, and shape irregularities.

### 6.4.1 Amplitude fluctuations: Time-walk correction

The discriminator following the preamplifier has a threshold setting that is aimed at avoiding false triggers; typically with a setting of 4-5 times the noise level. Pulses with different amplitudes cross the threshold at different times, large pulses being earlier than small pulses, creating the sensitivity to Landau amplitude variations. The simplest correction for this effect is to measure a quantity proportional to the pulse height and make a correction, pulse-by-pulse. The two most common solutions, illustrated in Figure 36, are Constant Fraction Discrimination (CFD) and Time-over-Threshold (ToT). There is also a third option that is very powerful and it is now starting to be used more widely called Multiple Samplings (MS). In this technique the signal is sampled multiple times, and a fit is used to define the particle time. CFD and ToT are simpler solutions, and they can be implemented per pixel within the read-out chip. MS is instead a rather complex algorithm as it requires the full digitization of the signal: this solution gives the best performance, but it can be used only for systems with a limited number of pixels as it needs a fair amount of computing power.

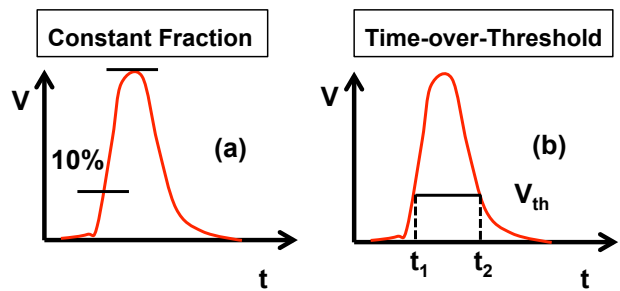

Figure 36 Basic circuits to correct for signal amplitude fluctuations: (a) Constant Fraction and (b) Time over Threshold. Figure taken with permission from [40].

Constant Fraction Discrimination sets the time of arrival of a particle when the signal reaches a given fraction of the total amplitude. This strategy involves defining the time of arrival of a particle ($t_{CF}$) based on the time the pulse reaches a given fraction of the maximum peak height $V_{Max}$, so it depends uniquely on the rising part of the pulse and not the tail. If the time for the pulse can be described as $t = g(\frac{V(t)}{V_{Max}})$, for any function g, then choosing a fraction, f = V(t)/$V_{Max}$, and recording the time the pulse reaches this fraction results in $t_{CF}$ = g(f), which has eliminated the dependence on the pulse height. An advantage of the constant fraction method is that the information is available very quickly if we use a constant-fraction-discriminator and that it doesn't need extra corrections.

Time-over-Threshold uses two time points to evaluate the arrival time of the particle ($t_{ToT}$) by applying a time-over-threshold ($t_2$-$t_1$) correction to the first time point $t_1$. This method requires therefore measuring the time duration the preamplifier signal is above the set threshold and then using this ToT value to correct $t_1$ with a formula optimized for the given electronics. This strategy requires the use of some additional logic such as an FPGA as the correction is calculated after the $t_1$ and ($t_2$-$t_1$) have been measured and recorded. Contrary to CFD, this technique requires measuring accurately both the rising and falling edge of the signal.

As CSA and BBA shape the input signal differently, the effectiveness of CFD and ToT techniques is different for the two types of amplifiers: both ToT and CFD can be used with CSA since amplitude and width of the output signal are proportional to the input charge, while with BBA CFD gives more sensitivity since the output signal width is almost constant making it difficult to apply the ToT technique.

CFD techniques work best if the pulse has the scaling property assumed to reasonable accuracy. It is very well suited to the Ultra-Fast sensors where the time for the peak of the pulse is very well defined, determined by the sensor thickness and saturated electron drift velocity in the sensor and independent of Landau fluctuations and the exact value of the gain in the sensor.

A combination of the methods is also attractive as it provides a quick time measurement from the sensor and a later careful correction and validation of the measurement. This however then requires more circuitry, which however is required in any case for the constant-fraction method that requires an arming discriminator to make sure the pulse is large enough in addition to the Constant-Fraction Discriminator.

### 6.4.2 Shape irregularities

The two amplifier types also employ different strategies to limit the impact of the non-uniform creation of e-h pairs. For current amplifiers the goal is to measure the time as close to the start of the pulse as practical, limited by the SNR. This is then sensitive to fluctuations at the start of the pulse rather than over the full collection period. Charge Sensitive Amplifiers, by integrating over the pulse, lump all the collected charge together and the Landau fluctuations create variations in the total charge collected, which has to be corrected for. The implementations of these strategies then are related to the next stages in the circuit, which are the discriminator steps.

## 6.5 Summary

Excellent timing and position resolution can be achieved only by a careful combination of sensors with electronics. Thin UFSD sensors provide fast and large current signals with a well defined shape, and the amplifiers need to take full advantage of this fact. For optimum performance using broad-band amplifiers, the time constant produced by sensor capacitance and preamplifier input impedance should be kept of the same magnitude of the current rise time, which places severe constrains for sensors with large capacitance. Time walk requires a necessary correction, but its effect can be kept below both the jitter and the irreducible contribution from charge non-uniformity.

# 7 Simulation of UFSD performances

Exploiting the prediction capabilities of WF2, we can gain insight into the timing performance of UFSDs and the effects of various geometries and different values of gain. In WF2, the simulated current is convoluted with the effects of the sensor capacitance and that of the chosen read-out system. As mentioned in section 4, WF2 can simulate the response of two types of amplifier: BBA, with selectable input impedance, gain, bandwidth and noise level, and CSA, with selectable rise and fall time, input impedance and trans-impedance. The user sets the noise values in WF2 accordingly to the set-up under test. It is important to stress that the numbers obtained in the simulation presented here are specific to the input parameters, however the trends and overall performances are general.

## 7.1 The effect of jitter and Landau noise on the UFSD time resolution

Figure 37 shows one of our main results: the improvement of the time resolution in thinner sensors. As summarized in section 3.6, jitter and time walk are minimized by having a higher slew rate, while

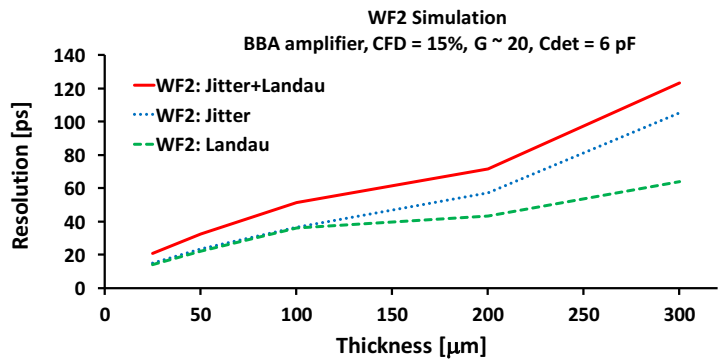

Figure 37 Jitter and Landau noise contribution to the time resolution as a function of the detector thickness

Landau noise, i.e. the effects due to a non-uniform creation of e-h pairs along the trajectory of the impinging particle, does not have an analytic expression and needs to be evaluated by simulation. In this example the capacitance, being rather small, does not slow down the signal rise time allowing the BBA to exploit the very fast signal rising edge. Simulation indicates that a 50-micron thick UFSD sensor has the capability of obtaining time resolutions between 30 and 40 picoseconds, for gain in the range 10-20. Similar results can be obtained using a CSA with short integration, provided that the detector capacitance is small enough (< 5 pF).

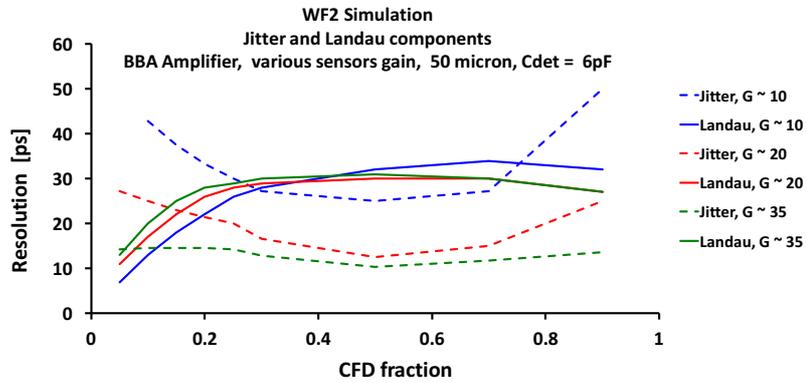

Figure 38 Jitter and Landau noise contributions to the total time resolution as a function of the CFD value.

Figure 37 shows the evaluation of the jitter and Landau noise components as a function of the sensor thickness for a UFSD of gain ~ 20. Interestingly, jitter and Landau noise contribute almost equally to the final values of time resolution, at least for sensor thickness below 150 micron. This prediction can be tested experimentally by comparing the time resolution in beam tests, where both components are present, to that of laboratory measurement obtained using very fast (a few picosecond) collimated 1064 nm laser shots, where only the jitter contribution is present.

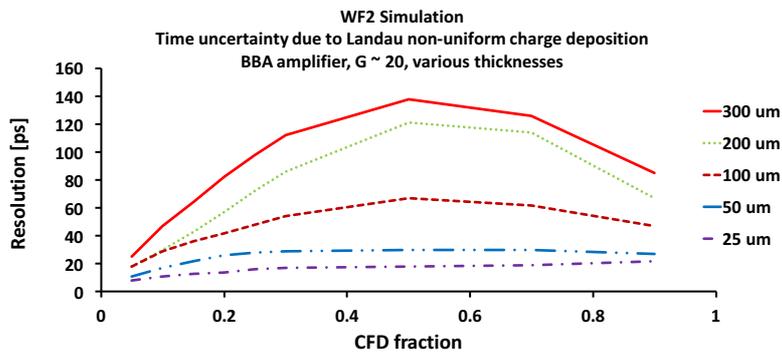

Figure 39 Landau noise contribution to the time resolution as a function of the CFD value for different detector thicknesses.

Figure 38 offers more details on the jitter and Landau noise contributions to the time resolution for a 50-micron thick sensor with different values of gain as a function of CFD settings. As expected, the jitter contribution follows the inverse of the signal derivative: at the start and at the end of the pulse the contribution increases, as the pulse shape is less steep. The effect of gain is rather predictable: higher gains yield to lower jitter. The Landau noise term, on the other hand, is rather insensitive to the gain value, but shows a clear dependence upon the CFD settings: it is minimized using the minimum possible threshold.

The dependence of the Landau noise on the sensor thickness as a function of the CFD values is shown in Figure 39: if from noise considerations the CFD threshold cannot be set low, for thick sensors the Landau noise is likely to represent the largest contribution to the time uncertainty [41]

## 7.2   Interplay of gain layer doping and bias voltage on the UFSD time resolution

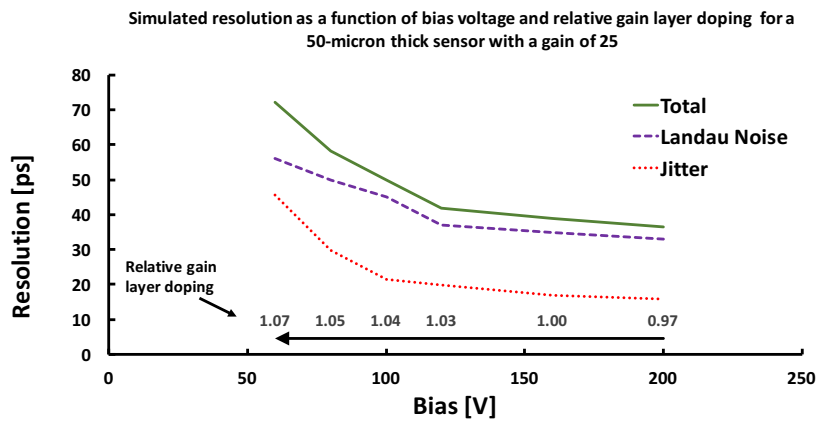

Figure 40 Time resolution of different combinations of gain layer doping and bias voltage for a 50-micron thick sensor with a constant gain of ~ 25

As discussed in section 2.3.2, the field in the gain region is the sum of the contributions from the doping of the gain layer and that from the external bias voltage: a given value of gain can therefore be achieved with various combinations of gain layer doping and bias voltage.  The impact on the time resolution of different combinations is shown in Figure 40 for a 50-micron thick sensor with a constant gain of ~ 25 where the values of doping are relative to the one used at $V_{bias}$ = 160 V. For high values of relative doping, 1.04 - 1.07, the gain of 25 is obtained at low values of bias voltage: under this condition, the drift velocity is not saturated and the time resolution is rather poor. This plot therefore corroborates the claim that saturated drift velocity is necessary to obtain good time resolution.

## 7.3   Summary

WF2 allows obtaining a deep insight into the mechanisms determining the time resolution.

Very good performances are obtained by a concurrent combination of optimized geometry, small capacitance, low gain and high electric fields. Simulation indicates that non-uniform charge deposition represents the most important effect in determining the time resolution, as the jitter contribution can be minimized by a gain value in the range 10 - 20. The results presented above suggest the conservative possibility to build a silicon tracker system with a time resolution of ~ 30 ps per plane.

# 8 Measurement of UFSD performance

## 8.1 UFSD productions

The LGAD technology was proposed and developed by the Centro Nacional de Microelectrónica (CNM) Barcelona, supported by national funding. The first publication containing measurements of LGAD sensors was presented in 2014 by CNM [11] while the first production of thin UFSD (50 μm) by CNM was presented in 2016 [42]. First beam test results on thin UFSD manufactured by CNM have been obtained in 2016 [43]. The Fondazione Bruno Kessler (FBK) has also designed [30] and produced LGAD sensors, up to now only 300-micron thick; first FBK production of thin LGAD is expected in early 2017. In the past 3 years CNM has manufactured a variety of LGAD designs, exploring different substrates (float zone (FZ), silicon-on-insulator (SoI), epitaxial (epi) with high and medium resistivity), reaching a well-controlled manufacturing capability. FBK has manufactured a single run of very high quality, exploring traditional LGAD design, segmented p-side read-out and AC coupling read-out. First results on LGAD sensors manufactured by Hamamatsu Photonics have been shown at the TREDI 2017[9] conference.

---

[9] http://tredi2017.fbk.eu/

## 8.2   The multiplication mechanism

The key ingredient to UFSD performance is the multiplication mechanism, which should ensure a controlled, "low gain" value and minimum multiplication noise, obtained by tailoring the field in such a way that the electrons cause multiplications but the holes do not. The "low gain" requirement is actually quite challenging as it implies controlling the doping layer concentration to a few per cent: Figure 41 shows the gain value measured in 3 Hamamatsu 50-micron thick LGAD with different levels of doping concentration of the gain layer as a function of the detector bias voltage. As the picture shows, the range of doping concentration to achieve controlled low gain is rather limited.

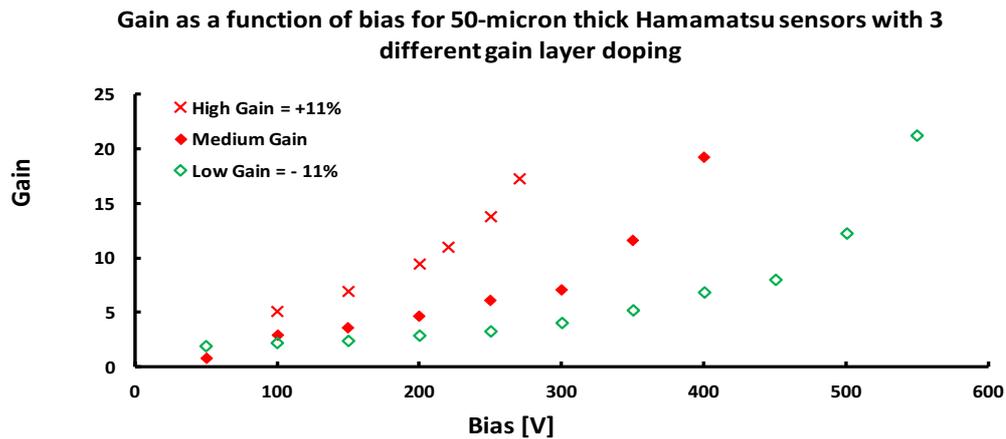

Figure 41 Gain values for 3 different implant doses as a function of detector bias voltage for 50-micron thick sensors produced by HPK.

### 8.2.1 Measuring the gain in LGADs

The gain in LGADs can be measured with a variety of methods by determining both the initial number of particles and the total number of particles after multiplication [44]. Often the gain of the LGAD is calculated from a comparison with an identical sensor without the special $p^+$ implant or from the known energy loss of MIPs. On the other hand, injection of electrons from the back side into the $p^{++}$ substrate either with a red laser or α-particles offers a self-calibrating measurement since it creates a very localized electron cloud which at first, while drifting towards the junction, creates the initial current signal, and then, when entering the gain layer, generates an equally localized hole cloud drifting back to the $p^{++}$ substrate as shown in Figure 42.

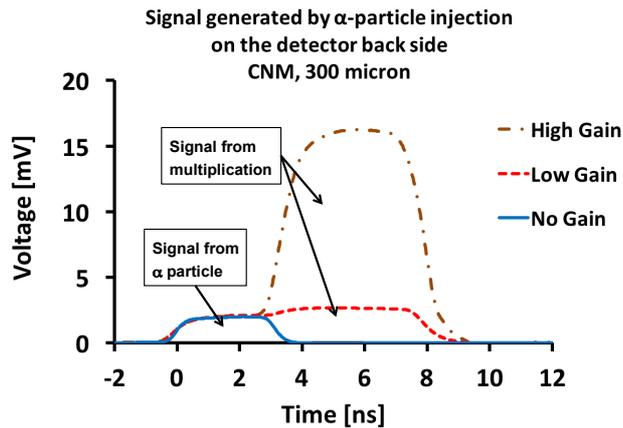

Figure 42 Self-calibrating gain determinations with backside electron injection using α-particles. The gain is defined as the ratio between the total area and the area of the initial e⁻ pulse. Figure taken with permission from [12].

### 8.2.2 Voltage dependence of gain and current

The field strength in the gain layer depends both on the doping concentration of the $p^+$ gain layer and the bias voltage: since the largest fraction of the electric field is generated by the $p^+$ implant, the dependence of the gain on the bias voltage is quite gentle, allowing having gain in a large interval of bias voltage values. Figure 43 shows how the gain changes as a function of the detector bias voltage for a 50 micron and two 300-micron thick sensors: in 300-micron sensors it takes 6 times the amount of external V bias to obtain the same change of electric field strength. Even for a 50 micron LGAD sensor the gain varies quite smoothly in the interval 50-150V. This capability is in sharp contrast with the operation of Silicon photomultipliers where the working voltage interval is at most 2-3V.

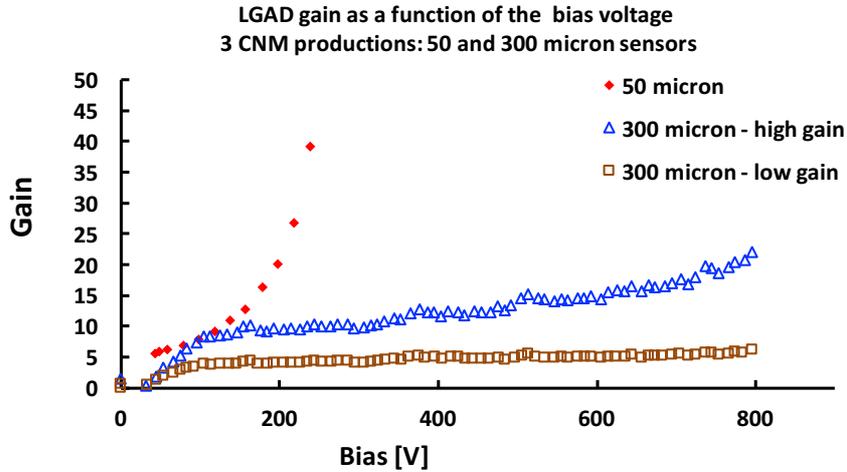

Figure 43 Gain variation as a function of the sensor bias voltage for 50 μm and 300 μm thick sensors.

The exponential dependence of the gain on the electric field can be extracted from signals generated by charged particles and from the leakage current, since in both cases the drift of charges is similar. As the measurements shown in Figure 44 indicate, this is indeed true in a region where the gain is below 20. Above that gain value both gain and current increase more steeply.

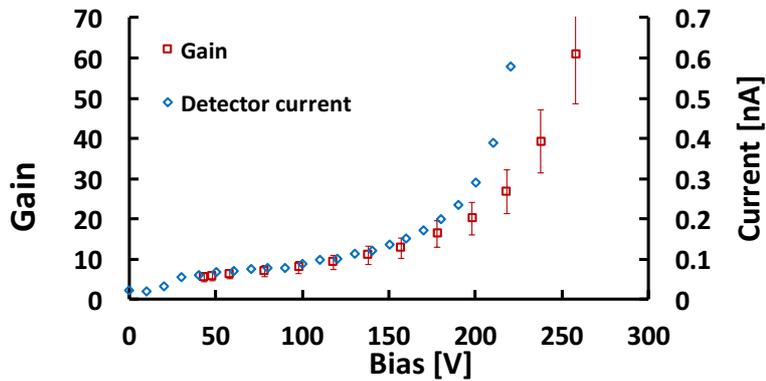

Figure 44 Gain and current dependence as a function of the sensor bias voltage. Figure taken with permission from [43].

### 8.2.3    Response to a mono-energetic charge deposition

The multiplication mechanism has been studied using mono-energetic light impulses from a 1064 nm pico-laser system calibrated to release the same amount of energy as that of a MIP. The laser shots allow studying the multiplication mechanism without the added complication of non-uniform charge distributions. Splitting the laser light into two detectors and using one to normalize the other was used to remove the laser amplitude jitter. As equation (8-1) shows, two components are contributing to the spread of the amplitude A: the electronics jitter and the excess noise factor:

$$\left(\frac{\sigma_{Tot}}{A}\right)^2 = \left(\frac{\sigma_{jitter}}{A}\right)^2 + \ \text{k} \cdot (G^x)^2, \qquad (8\text{-}1)$$

where k is a multiplicative constant.

The results show that the resolution is dominated at low gain values by the electronic term while at higher gain a contribution from the excess noise factor becomes visible with a value in line with simulation.

### 8.2.4    Temperature effects

The operating temperature has a strong impact on the UFSD performances: as the temperature

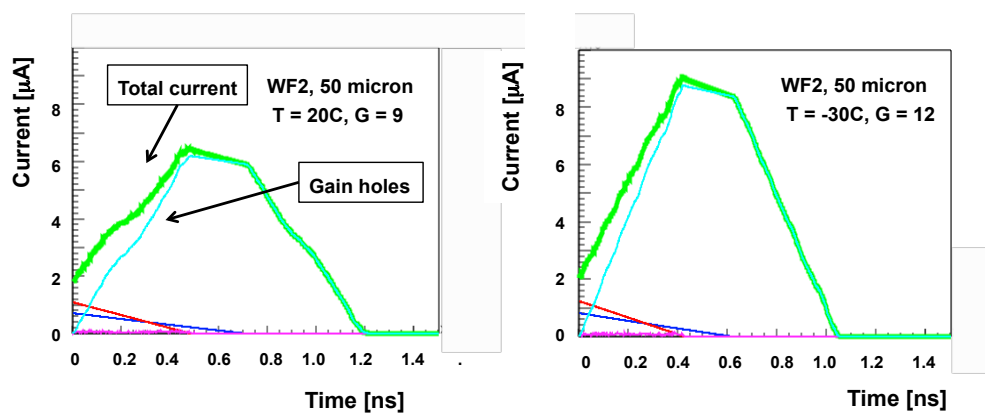

Figure 45 Temperature effects on UFSD performances: the drift time decreases and the gain increases. Left side T = 20 °C, gain = 7, right side: T = -30 °C, gain = 12 (WF2 simulations)

decreases the gain increases (see section 4.1), the drift velocity increases and the sensor leakage current decreases. The first two effects are shown in Figure 45, according to the Massey model [27] in the WF2 simulation: in addition to the total signal current being higher going from 300 K to 250 K as the gain increases from G = 9 to G = 12, the rise time becomes shorter due to the increased drift velocity.

As the temperature decreases, the breakdown voltage moves to lower sensor bias values, as shown for a 300 micron UFSD sensor on the left side of Figure 46 [23]. This fact causes a "gain-dependent" change of the gain value with temperature: the closer the starting gain value is to the breakdown voltage the larger will be the gain increase lowering the temperature.

The right side of the figure shows this effect for a 50 micron sensor for 3 bias voltages as a function of temperature: at a bias voltage of $V_{bias}$ = 210 V the change in gain is more than a factor of two than that at $V_{bias}$ = 150 V.

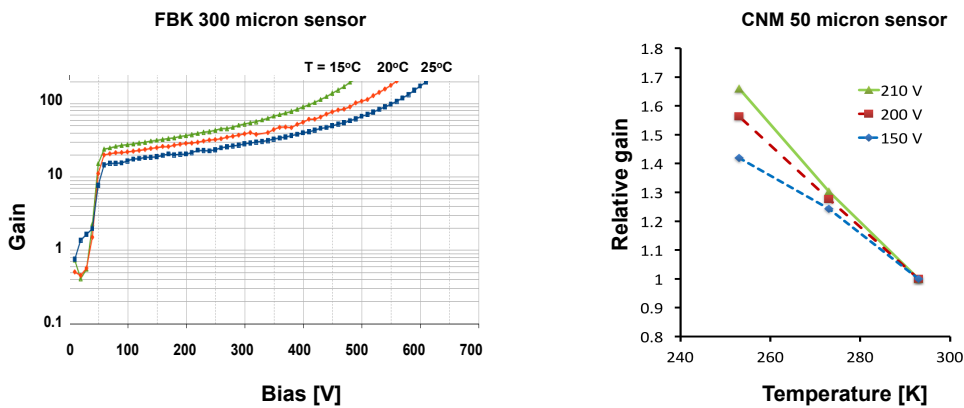

Figure 46: Left side: gain dependence as a function of the sensor bias voltage for 3 different temperatures. Right side: gain dependence as a function of the temperature for 3 different bias voltages.

We stress that it is important to model and measure the temperature dependence of the gain for each specific doping profile, as the change depends on the specific doping of the gain layer. A gain too high at room temperature will result in an early breakdown when operating the sensors cold, and will prevent using the sensors with a bias voltage high enough to saturate the drift velocity.

## 8.3 Results on time resolution

### 8.3.1 Beam test results from 50-micron thick UFSD

In this section we report on the results obtained in a beam test at CERN with π-mesons with a momentum of 180 GeV/c. Several 50-micron thick 1.2x1.2 mm$^2$ UFSD sensors produced by CNM were used with BBA amplifiers and a trigger board comprising of a SiPM coupled to a quartz bar [43]. This beam test, coupled with complementary laser measurements performed in our laboratories, provided the opportunity to perform detailed studies of the mechanisms governing UFSD time resolution and to compare these measurements to the simulation.

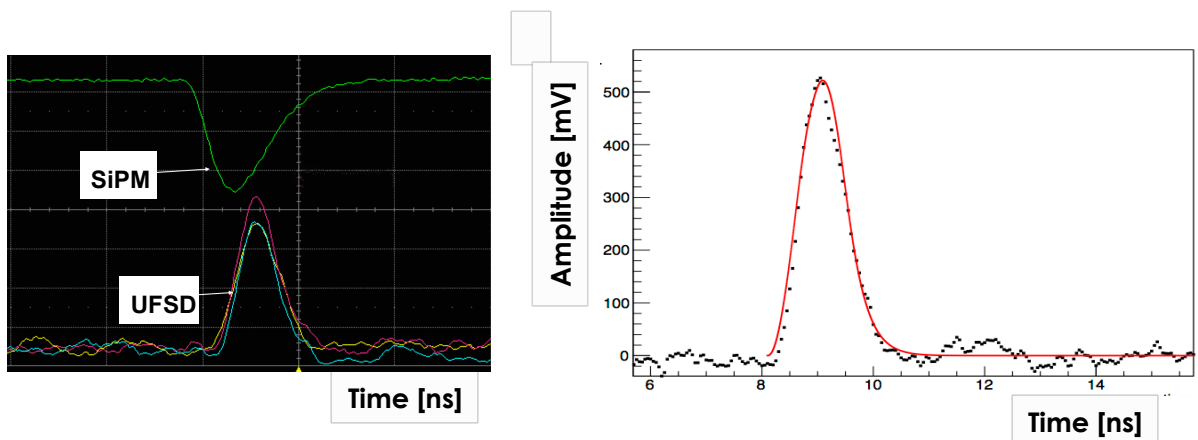

Figure 47  Left side: Signals of a beam test event showing the coincidence of 3 50-micron thick UFSD sensors and the SiPM trigger counter. Right side: Data – WF2 simulation  (solid line). Figure taken with permission from [43].

The left side of Figure 47 shows typical beam test signals, and the right side a comparison between data and WF2, demonstrating the capability of WF2 to reproduce the UFSD signals accurately. The signals are very fast, with low noise and large slew rate, ideal for timing studies. The time resolution of each sensor and that of the SiPM has then been obtained from the time differences between pairs of UFSD and between each UFSD and the SiPM, yielding two values for each UFSD. The time resolution of combined UFSD has been evaluated as the difference between the average time of two or three UFSD and the SiPM, Table 3.

Table 3 Timing resolution for single (N = 1), doublet  (N = 2) and triplets (N = 3) of UFSD at bias voltages of 200V and 240V. Table taken with permission from [43].

| UFSD Timing resolution | | |
|---|---|---|
| | **Vbias  = 200 V** | **Vbias = 230 V** |
| **N = 1** | 34 ps | 27 ps |
| **N = 2** | 24 ps | 20 ps |
| **N = 3** | 20 ps | 16 ps |

The results of Table 3 agree well with the expected σ(N) = $1/\sqrt{N}$ behaviour, demonstrating that the 3 sensors are of equal high quality. The timing resolution of a single UFSD is measured to be 34 ps for 200 V bias and 27 ps for 240 V bias. A system of three UFSD has a measured timing resolution of 20 ps for a bias of 200 V, and 16 ps for a bias of 240 V. The time resolution of a single UFSD is measured to decrease with increased gain G like G$^{-0.36}$.

### 8.3.2   Jitter and Landau noise contributions to the time resolution

As explained in sections 3.2 and 4.2, amplification distortion, jitter and non-uniform charge deposition contribute to the system time resolution.  In order to disentangle the effect of non-uniform charge deposition from the other effects, a comparison between beam test and laser time resolution measured with the same read-out chain is shown in Figure 48.

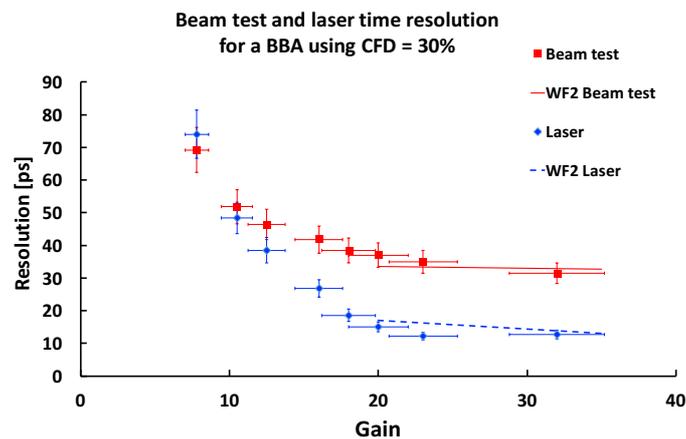

Figure 48 Comparison between beam test and laser time resolutions as a function of gain.

The laser data are affected only by the jitter and amplification distortion therefore the difference between the two sets of points is a direct measurement of the effect of non-uniform charge deposition on the time resolution. Superposed to the points are the predictions from the WF2 simulation, showing an excellent agreement. Figure 48 indicates that when the gain is above ~10, the slew rate is such that the electronic contribution is becoming less relevant and that the most important contribution to the time resolution comes from non-uniform charge deposition.

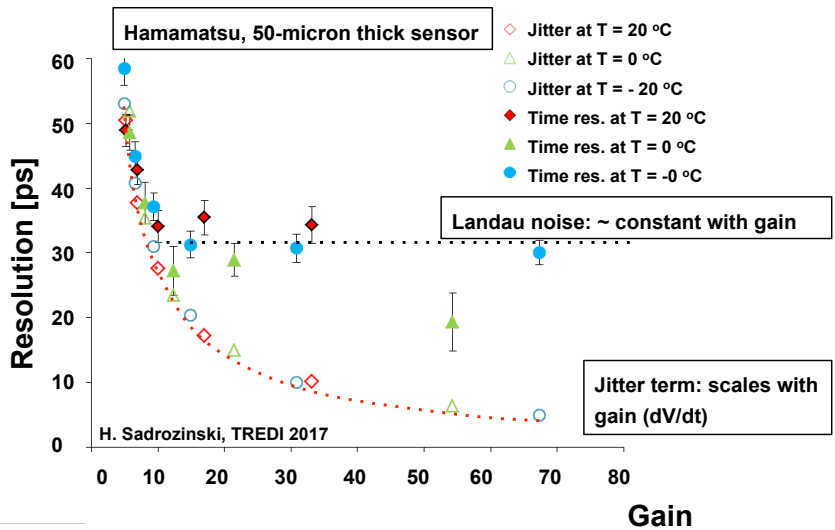

Figure 49 Jitter and the total time resolution as a function of gain for 50-micron thick Hamamatsu sensors for 3 different temperatures. Plot from [46].

Figure 49 shows the jitter and the total time resolution as a function of gain for 50-micron thick Hamamatsu sensors for 3 different temperatures, confirming the saturation of the time resolution at high gain. As the gain increases, the jitter decreases while the effect of non uniform charge deposition, that does not depend strongly on the gain value, provides a constant ~ 30 ps contribution to time resolution. This plot also clearly shows the idea behind UFSD: use the minimum gain that allows reaching the minimum value in time resolution, in this case a value G = 15.

Figure 50, combining laser and beam test results, explores the effect of the comparator threshold on the time resolution. At low values of CFD, the slope of the signal is not at its maximum, and the jitter contribution is still relevant, while when the slew rate increases the dominant contribution is given by non-uniform charge distribution.

The right side of Figure 50 complements the left side by showing how the time resolution changes as a function of CFD value for 3 different gain values: as the gain increases, there is a global shift towards lower time resolutions, with the jitter and non-uniform contributions interplaying to give an almost flat dependence of the time resolution on the CFD value.

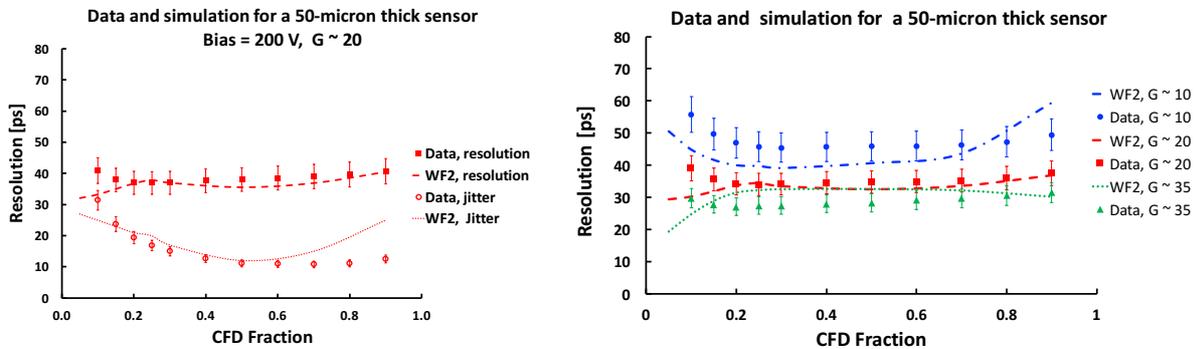

Figure 50 Time resolution of a 50-micron thick UFSD sensor. Left side: The total resolution and the jitter contribution as a function of CFD value. Right side: Time resolution as a function of CFD threshold for 3 different gain values.

### 8.3.3   Effect of multiple sampling on the time resolution

An obvious way to improve on the single CFD measurement is to use multiple CFD on the pulse and average the results: unfortunately this technique does not work due to very strong correlations between points. Figure 51 shows the measured time resolution obtained averaging two points on the rising edge of the signal for beam test and laser data as a function of the distance between the two points. The beam test data points show a perfect correlation, rendering the use of any type of multiple thresholds useless. The laser data, on the other hand, are less correlated, showing that as the distance between the two points increases, the time resolution improves.

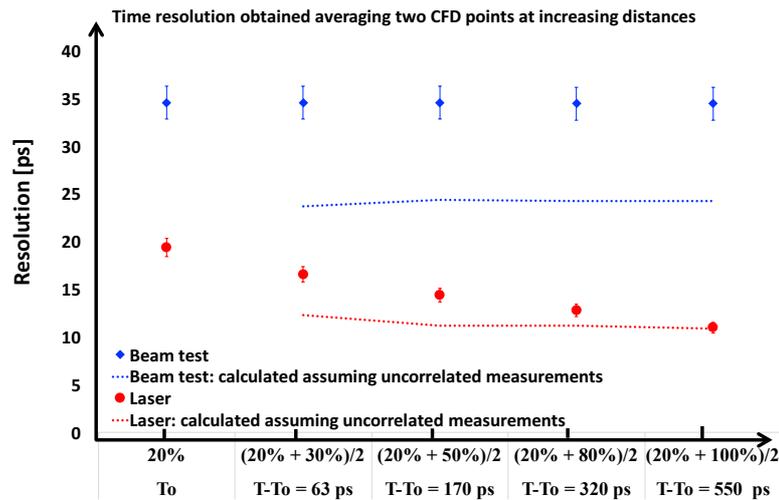

Figure 51 Time resolution obtained averaging two points at increasing distance in time. Beam test data show that the use of multiple CFD does not improve the resolution due to the strong correlation among points.

### 8.3.4 Effect of sensor geometry on the time resolution

In the past few years several productions of LGAD sensors have been manufactured by CNM and FBK exploring a large range of construction possibilities. Sensors with many different geometries, gain layer and bulk doping, bulk materials and thickness have been available for testing. Figure 52 shows a condensed summary of the timing studies together with a comparison of predicted results from simulation for a subset of geometries [13]. The results for 300-micron thick sensors elucidate the influence of gain and capacitance on the time resolution: smaller capacitances and higher gains yield to better performances. As an example, the simulation of one particular 300-micron geometry is also shown on the plot (red empty square).

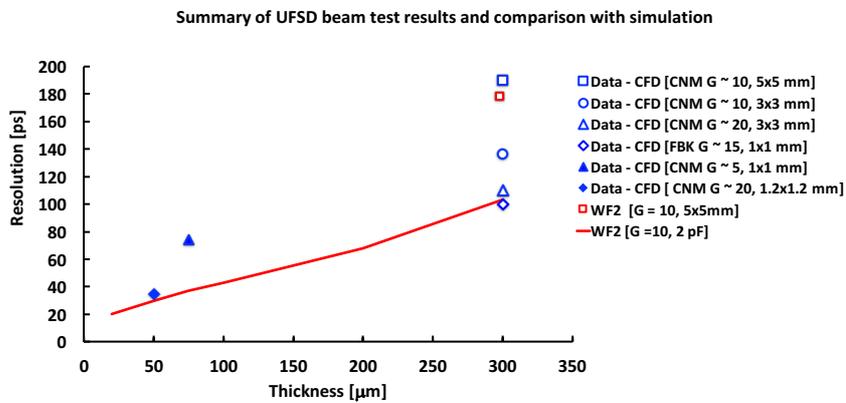

Figure 52 Summary the time resolutions achieved with different type of UFSD sensors.

The solid line shows how, according to simulation, the time resolution improves for a UFSD with a capacitance of 2pF and a gain of 10, suggesting a time resolution of about 30-35 ps for a 50-micron thick sensor. Two thin UFSD productions from CNM have been measured at a beam test: 75 micron with low gain, and 50 micron, where a time resolution of $\sigma_t = 34\ ps$ was reached at a bias voltage of 200V and a gain of ~20 [43]. The data were collected with the LGAD connected to a BBA with input impedance of about 25 Ohms.

## 8.4   Measurements of the properties of irradiated sensors

### 8.4.1   Bulk and gain layer contributions to the gain value in irradiated sensors

As it is shown in Figure 32, the key to maintain a constant gain value at high particle fluences it is to increase the bias voltage. This is possible only in thin sensors (50-micron thick or less) where the external voltage can create an electric field of the order of 250-300 kV/cm without causing electrical breakdown in other areas of the detector. Figure 53 shows the bias voltage necessary to maintain a gain G = 10 as a function of irradiation for 50-micron thick CNM sensors implanted with a shallow gain layer, using the irradiation results from [45] (squares), the initial acceptor removal parameterisation of equation (5-4) with the value of the parameters shown in Figure 27, and the Massey model of impact ionization [27], The very good agreement between the WF2 predicted values and the measured points indicates the correctness of the initial acceptor removal model and the appropriate gain parameterisation of the Massey equations in this range of the electric field. The thick solid line indicates the bulk contribution to the total gain value (right y-axis): it starts to be important at ~1e15 $n_{eq}$/cm$^2$, and it becomes ~ 50% at ~2e15 $n_{eq}$/cm$^2$. It is therefore key to operation in high radiation environments that the sensors after irradiation can sustain a biasing voltage high enough to assure a gain value above 10.

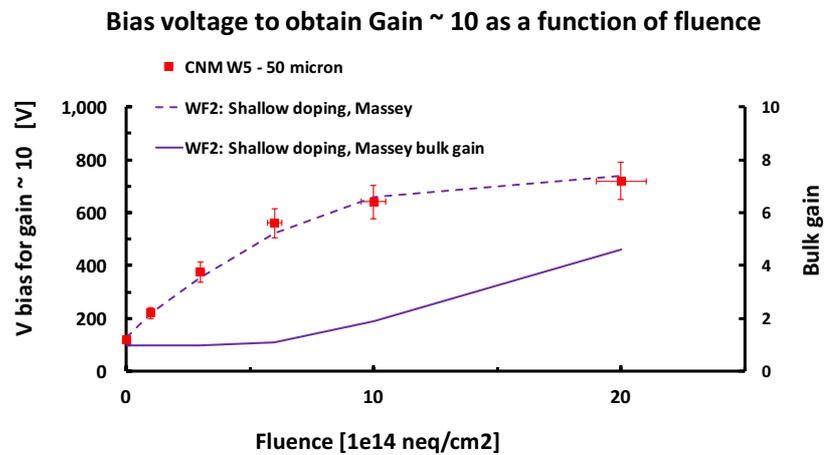

Figure 53 Necessary bias voltages to maintain a gain = 10 as a function of irradiation for 50-micron thick CNM sensors implanted with a shallow gain layer. Plot from 0.

Figure 54 shows the value of gain as a function of the bias voltage for three different CNM sensors 50-micron thick, two with the same initial doping of the gain layer but irradiated at different fluences ($\Phi$ = 6e14 $n_{eq}$/cm$^2$ and $\Phi$ = 2e15 $n_{eq}$/cm$^2$, respectively), and one with the initial doping of the gain layer 5% higher, irradiated at $\Phi$ = 6e14 $n_{eq}$/cm$^2$. The plot clearly shows the strong influence of the initial doping level on the performances after irradiation, indicating that a rather small initial difference (5%) has important consequences after irradiation. It is important to stress, however, that in this case higher level of initial gain layer doping are actually detrimental to the performance before irradiation (see section 7.2) since to avoid electrical breakdown the bias voltage needs to be kept below the values required to saturate the drift velocity. It is important to note that the simulations reported in Figure 32 on gain restoration are in very good agreement with the measurements shown in Figure 54.

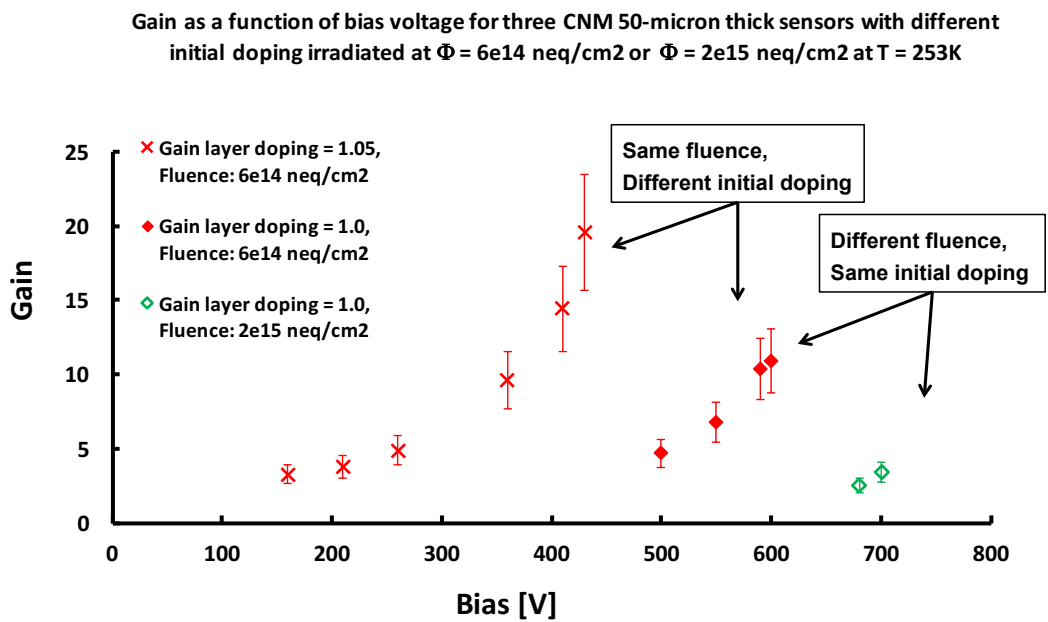

Figure 54 Value of gain as a function of the bias voltage for three different CNM sensors 50-micron thick, two with the same initial doping of the gain layer but irradiated at different fluences ($\Phi$ = 6e14 $n_{eq}$/cm$^2$ and ($\Phi$ = 2e15 $n_{eq}$/cm$^2$), and one with the initial doping of the gain layer 5% higher. Error bars reflect a common gain uncertainty of 20%.

### 8.4.2    Time resolution in irradiated sensors

The time resolution of irradiated detectors might deteriorate for several effects, most notably the increase in noise due to a larger leakage current, and a lower gain due to initial acceptor removal. Figure 55 shows the time resolution and jitter as a function of gain for three CNM 50-micron thick sensors with three different initial levels of gain layer doping (0.95, 1.0, and 1.05), irradiated at two

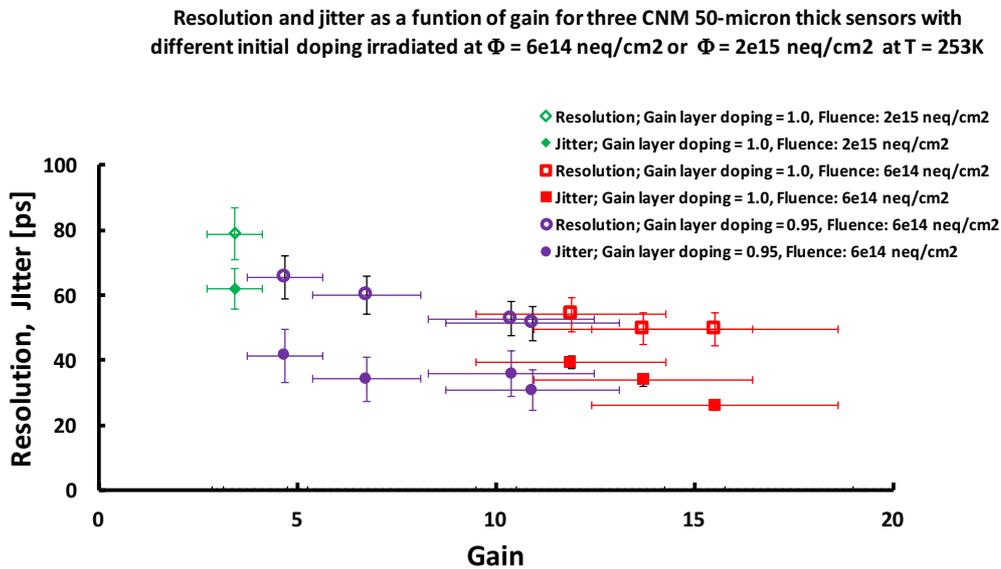

**Resolution and jitter as a funtion of gain for three CNM 50-micron thick sensors with different initial doping irradiated at Φ = 6e14 neq/cm2 or Φ = 2e15 neq/cm2 at T = 253K**

◇ Resolution; Gain layer doping = 1.0, Fluence: 2e15 neq/cm2
◆ Jitter; Gain layer doping = 1.0, Fluence: 2e15 neq/cm2
□ Resolution; Gain layer doping = 1.0, Fluence: 6e14 neq/cm2
■ Jitter; Gain layer doping = 1.0, Fluence: 6e14 neq/cm2
○ Resolution; Gain layer doping = 0.95, Fluence: 6e14 neq/cm2
● Jitter; Gain layer doping = 0.95, Fluence: 6e14 neq/cm2

Figure 55 The time resolution and jitter for three CNM 50-micron thick sensors irradiated at two fluences (6e14 $n_{eq}/cm^2$ and 2e15 $n_{eq}/cm^2$) as a function of gain. Plot from [46]. Horizontal error bars reflect a 20% common uncertainty on the gain determination.

fluences ($\Phi = 6e14$ $n_{eq}/cm^2$ and $\Phi = 2e15$ $n_{eq}/cm^2$). The plot shows the important feature that, even after irradiation, the most important parameter for the time resolution remains the gain value and that the jitter contribution, determined by the noise, is still rather small even after a fluence of 6e14 $n_{eq}/cm^2$.

## 8.5    Summary

Thin UFSD sensors allow obtaining excellent time resolution: we measured $\sigma_t = 34$ $ps$ for a gain of ~ 20 at a beam test with π-mesons with a momentum of 180 GeV/c, and $\sigma_t = 20$ $ps$ for the average of three UFSD. As the gain increases, the contribution from jitter becomes insignificant, and the time resolution is determined by non-uniform charge deposition. Using a very low constant fraction discriminator value to control this effect does not improve the time resolution appreciably since the jitter term remains relevant at the start of the signal. The shape of the signal is such the use of multiple measurements to improve time resolution does not work, due to the long-range correlation induced by the irregular charge deposition and the electronic noise. Irradiation with large fluences of hadrons causes deterioration of the time resolution mainly due to the loss of gain, yet in case a gain of $G > 10$ can be maintained by increasing the bias voltage, a time resolution of $\sigma_t = 40\ ps$ can be achieved. A concentrated investigation of the properties of UFSD after irradiation is taking place even while this article is being published. For an update the reader might refer to the presentations at the bi-annual RD50 Workshop [48].

# 9   Summary and outlook

We have evaluated the characteristics that silicon sensors should have in order to achieve excellent time resolution in addition to position resolution, exploiting the enhanced signals from silicon detectors with internal gain provided by the LGAD technology, and guided by an ad-hoc simulation program, Weightfield2 (WF2). Specifically we postulate that (i) saturated drift velocity of the charge carriers, (ii) uniform weighting field, (iii) low gain, (iv) small thickness, and (v) small pad volume be key elements of the sensor design. We called the sensors with this design Ultra-Fast Silicon Detectors (UFSD). Since UFSD have been produced successfully in the past few years at several silicon foundries, we were able test these requirements with actual devices, and found good agreement between simulations and data.

Radiation hardness studies have shown the possibilities to use UFSD up to fluences of $\Phi \sim 1e15$ $n_{eq}/cm^2$. A strong R&D program currently under way is exploring new UFSD designs to further increase the UFSD radiation tolerance.

In parallel to the sensor design, we carried on the development and production of dedicated read-out electronics, tailored to the output signal of UFSD: we propose the notion that the optimal front-end design for timing application has to be able to follow the fast rising edge of the signal, and not integrate the signal as previously done in most silicon detector read-out.

Combining UFSD sensors with our electronics we achieved at beam tests a time resolution below 30 ps, a result that positions silicon detectors among the best timing devices.

The next frontier of UFSD is their use in experiments that require large area coverage, either at HL-LHC or in other applications. This next step will require the production of large quantities of high quality UFSD, their associated electronics and the development of sophisticated clock distribution systems.

# 10  Acknowledgments

We thank our collaborators within RD50, ATLAS and CMS who participated in the development of UFSD. Our special thanks to the technical staff at UC Santa Cruz, INFN Torino, CNM Barcelona, FBK Trento and HPK Hamamatsu. This work was partially performed within the CERN RD50 collaboration.

The work was supported by the United States Department of Energy, grant DE-FG02-04ER41286. Part of this work has been financed by the European Union's Horizon 2020 Research and Innovation funding program, under Grant Agreement no. 654168 (AIDA-2020) and Grant Agreement no. 669529 (ERC UFSD669529), and by the Italian Ministero degli Affari Esteri and INFN Gruppo V. We thank dr. R. Arcidiacono and dr. M. Mandurrino for the critical reading of the manuscript.